\documentclass[journal, onecolumn, draftcls]{IEEEtran}

\usepackage{cite}
\usepackage{amsmath ,amsfonts}
\usepackage{bm}
\usepackage{nicefrac}
\usepackage{siunitx}
\usepackage{graphicx}
\usepackage{xcolor}
\usepackage{subcaption}
\usepackage{siunitx}
\DeclareSIUnit{\molar}{M}
\usepackage[version=4]{mhchem}    
\usepackage{subcaption}
\usepackage{acronym}
\usepackage{nicefrac}
\usepackage{mathtools}
\usepackage{multirow}
\usepackage{amssymb}
\usepackage{booktabs}
\usepackage{hyperref}
\hypersetup{hidelinks}
\acrodef{CAM}{chorioallantoic membrane}
\acrodef{ICG}{indocyanine green}
\acrodef{IoBNT}{Internet of BioNanoThings}
\acrodef{MC}{molecular communications}
\acrodef{DED}{days of embryonic development}
\acrodef{TX}{transmitter}
\acrodef{RX}{receiver}
\acrodef{nRX}{natural receiver}
\acrodef{sRX}{synthetic receiver}
\acrodef{ISI}{inter symbol interference}
\acrodef{ROI}{region of interest}
\acrodef{PBS}{particles-based simulations}
\acrodef{CIR}{channel impulse response}
\acrodef{RMSE}{root mean square error}

\newcommand{\veff}{v_\mathrm{eff}}

\newcommand{\Deff}{D_\mathrm{eff}}
\newcommand{\Leff}{L_\mathrm{eff}}

\newcommand{\Hquad}{\hspace{0.5em}} 
\newcommand\varcal[1]{\text{\usefont{OMS}{cmsy}{m}{n}#1}}

\begin{document}
\title{The CAM Model: An in vivo Testbed for Molecular Communication Systems}

\author{F. Vakilipoor, A. Ettner-Sitter, L. Brand, S. Lotter, 
        T. Aung,\\ S. Harteis, R. Schober, and M.~Sch\"afer
\thanks{F. Vakilipoor, L. Brand, S. Lotter, R. Schober, and M. Sch\"afer are with the Institute for Digital Communications, Friedrich-Alexander-Universität Erlangen-Nürnberg, Germany (fardad.vakilipoor@fau.de). A. Ettner-Sitter and S. H\"arteis are with the Institute for Molecular and Cellular Anatomy, University of Regensburg, Germany. Thiha Aung is with the Deggendorf Institute of Technology.} 
\thanks{This paper was presented in part at the 11th ACM International Conference on Nanoscale Computing and Communication~\cite{schafer2024chorioallantoic}.}
}

%


\maketitle

\begin{abstract}
Molecular communication (MC) research is increasingly focused on applications within the human body, such as health monitoring and drug delivery, which require testing in realistic and living environments. Thus, elevating experimental MC research to the next level requires developing realistic \textit{in vivo} experimental testbeds. In this paper, we introduce the chorioallantoic membrane (CAM) model as the first versatile 3D \textit{in vivo} MC testbed. 
The CAM itself is a highly vascularized membrane formed in fertilized chicken eggs and the CAM model has gained significance in various research fields, including bioengineering, cancer research, and drug development. Its versatility, reproducibility, and realistic biological properties make it perfectly suited for next-generation MC testbeds, facilitating the transition from proof-of-concept systems to practical applications. 
In this paper, we provide a comprehensive introduction to the CAM model, its properties, and its applications in experimental research. Additionally, we present a characterization of the CAM model as an MC system. As an experimental study, we investigate the distribution of fluorescent molecules in the closed-loop vascular system of the CAM model. 
We derive an analytical model based on the wrapped normal distribution to describe the propagation of particles in dispersive closed-loop systems, where the propagation of particles is mainly influenced by diffusion and flow. 
Based on this analytical model, we propose parametric models to approximate the particle propagation dynamics inside the CAM model. The model parameters are estimated via curve fitting to experimental results using a nonlinear least squares method. We provide a dataset containing experimental results for 69 regions in 25 eggs, on which we evaluate the proposed parametric models. Moreover, we discuss the estimated parameters, their relationships, and plausibility. Furthermore, we investigate and develop a parametric model for the long-term behavior of particles in the CAM model and their accumulation in the chick embryo’s liver.
\end{abstract}

\begin{IEEEkeywords}
Chorioallantoic Membrane, Molecular Communication, Wrapped Normal Distribution, Diffusion, Indocyanine Green, ICG Accumulation.
\end{IEEEkeywords}

%
\IEEEpeerreviewmaketitle

\section{Introduction}
\label{sec:intro}
Synthetic \ac{MC} is an emerging interdisciplinary research field at the intersection of life sciences and engineering \cite{Nakano2011}. \ac{MC} naturally occurs in biological systems and is envisioned to enable communication between synthetic nanomachines and biological entities. In recent years, numerous potential applications of MC have been identified across various sectors, including industry, environmental monitoring, and medicine \cite{Farsad2016,Felicetti2016}.
Despite the recent increase of experimental work, most \ac{MC} research remains theoretical. Bridging the gap between theoretical concepts and their practical applications is a significant challenge \cite{Lotter2023b}. 
In the medical sector, \ac{MC} often focuses on in-body applications, necessitating the validation of concepts and technologies in realistic \textit{in vivo} environments. However, many promising MC technologies are highly invasive and disruptive, making animal or human testing very difficult and currently almost impossible. 

\begin{figure}[t]
    \centering
    \includegraphics[width=0.6\linewidth]{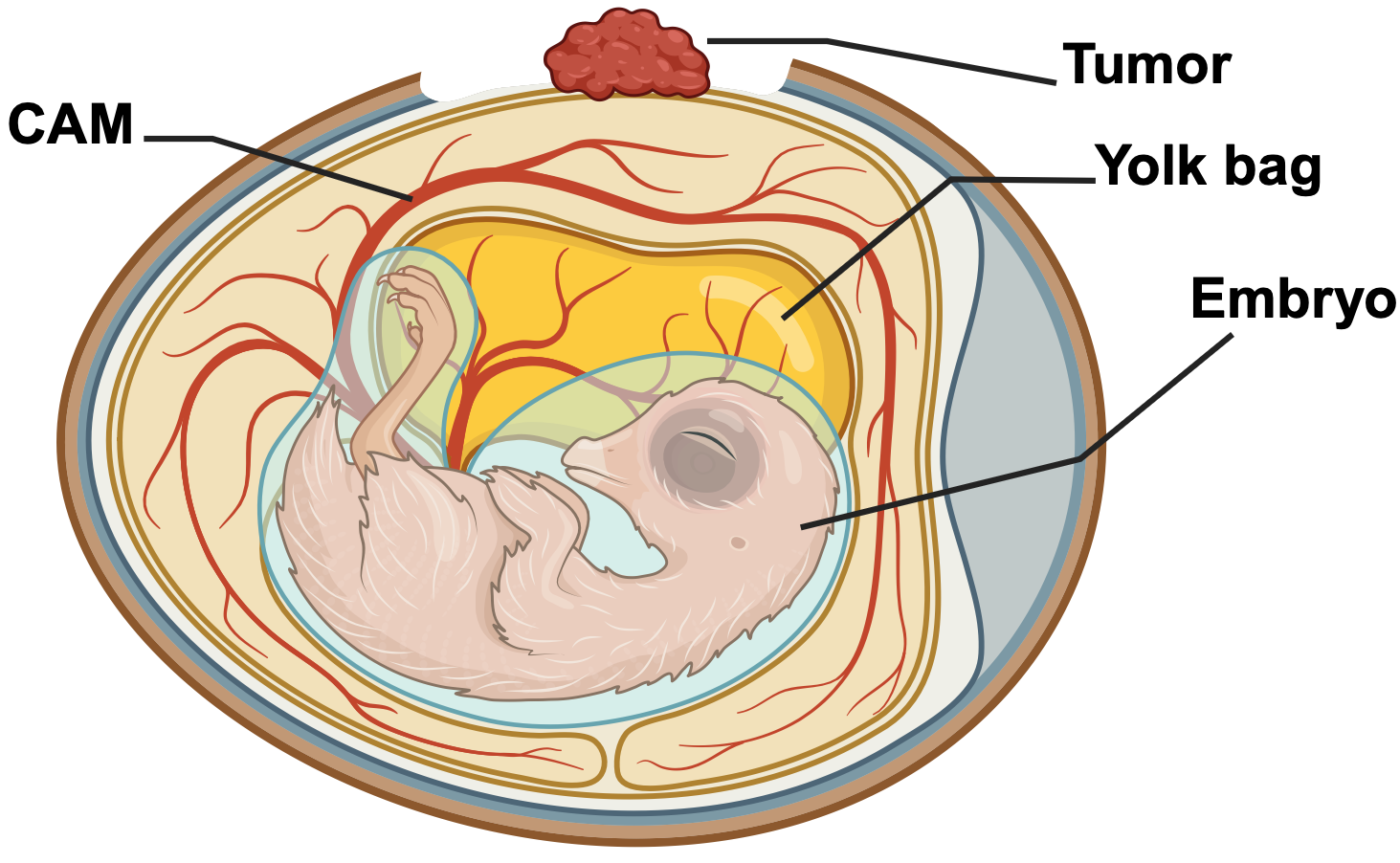}
    \caption{\small Schematic illustration of the CAM model consisting of the chicken embryo, the yolk bag, and the CAM. Furthermore, a tumor is engrafted on the CAM (Created with \url{BioRender.com}).}
    \label{fig:cam-schematic}
\end{figure}

Experimental work on in-body \ac{MC} is often focused on concepts for exchanging information between in-body devices, such as mobile nanosensors patrolling the cardiovascular system, and devices for the controlled delivery of drug molecules, and their experimental validation. Hereby, most testbeds established so far employ a tube-like channel (e.g., pipes or microfluidic chips) as a rough approximation of the cardiovascular system, where the signaling molecules are injected via a syringe or a valve \cite{Farsad2017,Unterweger2018,cali2024experimental,Pan2022,Wang2020,huang2024energy,lin2024testbed,Brand2022,Brand2024}. 
In~\cite{Unterweger2018}, a tube-based \ac{MC} testbed employing biocompatible magnetic nanoparticles was developed. This system featured an electronic pump acting as the \ac{TX}, alongside a second pump to ensure a consistent background flow for signal propagation. On the \ac{RX} side, a susceptometer was installed to generate electrical signals as the magnetic particles traversed through it.
In~\cite{cali2024experimental}, a prototype platform for molecule shift keying was proposed. The system comprised an infusion pump and valves as \ac{TX}, a tube-based microfluidic channel, and a compact charge-coupled device (CCD) as \ac{RX}. The \ac{RX} detected fluorescence signals emitted by graphene quantum dots (GQDs), which served as signaling particles.
In \cite{Pan2022}, a nanomaterial-based \ac{MC} testbed was proposed to mimic the cardiovascular system. Color pigments were employed as signaling molecules, which were detected by a color sensor as \ac{RX}. 
%
In~\cite{Wang2020}, a fluid-based testbed was reported employing a channel consisting of a network of tubes, two peristaltic pumps as \acp{TX}, and an electrical conductivity sensor as \ac{RX}.
%
%
A similar testbed was recently proposed in~\cite{huang2024energy}, where the information was encoded into the pH level of the fluid. 
%
The authors in \cite{Brand2024, Brand2024closed} proposed a closed-loop fluid-based testbed employing the green fluorescent protein dreiklang (GFPD) as a reversibly switchable biocompatible signal molecule. 
In \cite{lin2024testbed}, the authors recently introduced a fluid-based testbed using animal blood as propagation medium and flesh to replicate the damping effects of skin and tissue on signal detection.
As signaling molecule, biocompatible \ac{ICG} was employed and the concentration at the \ac{RX} was measured by an optical intensity sensor. The animal flesh was placed between the propagation channel and the sensor to mimic properties of human skin. 
For a more comprehensive overview on experimental \ac{MC} research, we refer to \cite{Lotter2023a,Lotter2023b}. 

All the testbeds in \cite{Farsad2017,Unterweger2018,cali2024experimental,Pan2022,Wang2020,huang2024energy,lin2024testbed,Brand2024closed,Brand2024} try to mimic realistic propagation environments. For example, studying blood as propagation environment \cite{lin2024testbed}, biocompatible signaling particles such as \ac{ICG} \cite{lin2024testbed}, GQDs \cite{cali2024experimental}, magnetic nanoparticles \cite{Unterweger2018} or GFPD \cite{Brand2024}, and closed-loop systems \cite{Brand2024closed}, is important for gaining insight into the operation of MC systems under real-world conditions. However, although these testbeds can successfully model some properties of real-world environments, provide a controllable setup, and allow the evaluation of the relevance of some parameters for \ac{MC} system design, they are still far away from realistic application environments such as the cardiovascular system. 
Therefore, despite this significant progresses in experimental \ac{MC} research in the last few years, the influence of realistic and living - \textit{in vivo} - environments on \ac{MC} system design has not been comprehensively investigated, yet. In order to enable the practical deployment of synthetic MC systems in the future, it is indispensable to elevate experimental \ac{MC} research to the next level by developing \textit{in vivo} testbeds. 
With this in mind, in this paper, we propose a 3D \textit{in vivo} \ac{MC} testbed based on the \acf{CAM} model, which provides an accessible model of a cardiovascular system including blood circulation and organs, see Fig.~\ref{fig:cam-schematic}.

The \ac{CAM} itself is a highly vascularized membrane which is found in the eggs of certain amniotes like birds and reptiles~\cite{gilbert2011developmental} and functions as a respiratory organ during embryo development \cite{Mapanao21, Valdes2002}. 
Moreover, human cells or tissues can be engrafted onto the \ac{CAM}, e.g., to study the effect of therapeutics. Over the years, the \ac{CAM} model has been established in different research fields and was originally used for angiogenic\footnote{Angiogenesis is the formation of new blood vessels~\cite{AngioRef}.} and anti-angiogenic therapeutic approaches. In recent years, however, it has gained importance in many different fields, such as bioengineering, transplant biology, cancer research, patient derived xenograft (PDX) research, and drug development \cite{Feder2020,kohl20223d,Wagner24}. As the \ac{CAM} model represents a simple and accessible \textit{in vivo} model, it is perfectly suited as next-generation \ac{MC} testbed. 
Therefore, in this paper, we propose the \ac{CAM} model as a versatile and realistic model of an \textit{in vivo} system for the validation and optimization of \ac{MC} technologies, to enable the transition from simple proof-of-concept MC systems towards practical applications \cite{Ettner2024,schafer2024chorioallantoic}. 
As a first step, to characterize the \ac{CAM} model as an \ac{MC} system, we investigate the particle distribution and accumulation inside the \ac{CAM} model's vascular system. To this end, we developed an experimental setup where we inject the fluorescent molecule \ac{ICG} into the \ac{CAM}'s vascular system and measure the \ac{ICG} intensity with a fluorescence camera. From the measurements it can be observed that injected particles distribute very fast inside the vascular system and after a steady state phase characterized by a uniform particle distribution, particles start to accumulate in the yolk bag and organs of the \ac{CAM} model. Based on this observation, we show that the particle distribution in the \ac{CAM} model can be divided into three phases, i.e., a \textit{transient}, a \textit{steady state}, and an \textit{accumulation phase}. To characterize the different distribution phases, we develop parametric models based on the wrapped normal distribution, capable of describing the particle distribution in highly-vascularized closed-loop systems. The parametric models are fitted to experimental measurement data by a nonlinear least squares method to demonstrate their validity and to show their capability of providing valuable insights regarding characteristic effects occurring in living environments. Moreover, the accuracy of the proposed parametric models is evaluated by the \ac{RMSE}.
%
%
%
The main contributions of this paper can be summarized as follows:
\begin{itemize}
    \setlength\itemsep{0em}
    \item For the first time, we propose the \ac{CAM} model as a versatile and accessible \textit{in vivo} \ac{MC} testbed, characterize its properties, and discuss its capabilities for the analysis and design of future \ac{MC} systems.

    \item We experimentally investigate the distribution of the fluorescent molecule \ac{ICG} inside the \ac{CAM} model. Our experiments show that particles quickly distribute uniformly in the \ac{CAM} model, before they accumulate in organs. We characterize the \ac{ICG} distribution in the \ac{CAM} model into three distinct phases, i.e., the \textit{transient phase}, representing the fast distribution in the vascular system, the \textit{steady state phase}, when particles are uniformly distributed inside the vascular system, and the \textit{accumulation phase}, when particles start to accumulate in the organs of the embryo. 

    \item We derive an analytical model based on a wrapped normal distribution, capable of describing the particle distribution in dispersive closed-loop systems. We validate the model by a comparison with results from \ac{PBS}. Inspired by the analytical model, we propose a basic parametric model and show that it is capable of approximating the transient and steady state particle distribution behavior in the \ac{CAM} model by fitting the model parameters to experimental measurement data.  

    \item We extend the basic parametric model to account for additional topological properties of vascularized networks, such as branching and the possible existence of multiple loops. We show that the extended model is in very good agreement with the experimental data for the transient and steady state distributions of particles inside the \ac{CAM} model. Moreover, we show that the physical model parameters obtained for the proposed model via fitting to experimental measurements are consistent and plausible.

    \item We present an initial experimental study on the accumulation of particles in the liver of the embryo, which is a natural clearance mechanism of the \ac{CAM} model. Moreover, we propose a first simple model to approximate the accumulation of particles in the liver and validate it by a comparison with experimental results. 

    \item We created an experimental dataset containing $69$ measurements in $25$ eggs on different \ac{DED}. The dataset contains measurements of the \ac{ICG} distribution inside the \ac{CAM} model and all parameter estimation results obtained for the developed parametric models. The dataset is shared with the community and will be continuously extended~\cite{ICGdistDataset2025}.
\end{itemize}

Compared to the conference version~\cite{schafer2024chorioallantoic} of this paper, we model the distribution of \ac{ICG} inside the \ac{CAM} model more carefully and divide it into three phases, i.e., the \textit{transient phase}, \textit{steady state phase}, and \textit{accumulation phase}. 
We extend the parametric model proposed for the molecule distribution in \cite{schafer2024chorioallantoic} to a mixture of wrapped normal distributions to capture the dynamics of the particle distribution in dispersive closed-loop systems. 
We demonstrate that the extended parametric model is capable of approximating the particle distribution in the \ac{CAM} model more accurately compared to the model proposed in~\cite{schafer2024chorioallantoic}. Moreover, we analyze the accuracy of the proposed parametric models based on the \ac{RMSE} between the results obtained for the proposed models and experimental measurements. Furthermore, we investigate the \ac{ICG} accumulation in the embryo's liver and present a corresponding simple parametric model.
Moreover, we created a dataset containing experimental measurement data of the transient and steady state distribution of \ac{ICG} inside the \ac{CAM} model~\cite{ICGdistDataset2025}.


The remainder of this paper is organized as follows: In Section~\ref{sec:CAM}, we introduce the \ac{CAM} model, its advantages and disadvantages, topology, and applications. In Section~\ref{sec:particleDis}, we characterize the \ac{CAM} model from an \ac{MC} point of view and propose an analytical model for the distribution of molecules in dispersive closed-loop systems. In Section~\ref{sec:est}, we introduce the proposed parametric models to capture the dynamics of the \ac{ICG} injection, distribution in the \ac{CAM} vascular system, and accumulation in the embryo's liver. Section~\ref{sec:exp} is divided into two parts. First, the proposed parametric models are employed to approximate the distribution of molecules inside the \ac{CAM} vascular system during the \textit{transient} and \textit{steady state phases} by fitting their parameters to experimental measurement data. Moreover, a comprehensive discussion of the estimated model parameters is provided, and the dataset is described in detail. Second, the accumulation of \ac{ICG} molecules in the embryo's liver during the \textit{accumulation phase} is studied and the proposed parametric model is used to approximate the dynamics of particle accumulation. In Section~\ref{sec:discuss}, we discuss the results obtained in this paper and propose several directions for future research. Our conclusions are presented in Section~\ref{sec:conc}, and the methodology for conducting the experiments is detailed in Appendix~\ref{sec:methods}.
\section{The Chorioallantoic Membrane Model}
\label{sec:CAM}
In this section, we provide a comprehensive overview on the \ac{CAM} model. We first provide a short introduction to the \ac{CAM} model, and discuss its advantages and limitations for experimental research. Then, we outline the main research directions within the \ac{CAM} literature.
Since the \ac{CAM} model has not been considered before in the \ac{MC} community, we provide an overview of its components, describe the main physical parameters of the developing vascular system, and explain the growing importance of the \ac{CAM} model in fields relevant for the MC community such as tumor and drug distribution research. 

\subsection{Introduction to the CAM Model}
The \ac{CAM} is formed in fertilized chicken\footnote{Although the CAM is present in many amniotes, including birds and reptiles, it is most commonly studied in fertilized chicken eggs. Due to their widespread availability and the extensive related literature, the term `CAM' is conventionally used to refer to the vascularized extraembryonic membrane in fertilized chicken eggs.} eggs as a highly vascularized extraembryonic membrane that functions as a respiratory organ during the embryo's development~\cite{Valdes2002}. 
For over a century, scientists have utilized the CAM model in medical research, as it can serve as a realistic model for \textit{in vivo} systems, bridging the gap between preclinical research and clinical trials \cite{Mapanao21}.
As early as $1911$, Rous and Murphy grafted chicken sarcoma cells onto the CAM model and observed tumor growth~\cite{rous1911tumor}. A year later, Murphy published the successful heterologous transplantation of tumors to the \ac{CAM}  model~\cite{murphy1912transplantability}. In the $1930$s, the CAM was exploited for the first time for the cultivation of viruses and bacteria~\cite{goodpasture1931cultivation,moore1942chorioallantoic}.
Later, the \ac{CAM} model was also used for vascular studies, such as blood vessel development and testing of pro-angiogenic and anti-angiogenic properties of substances~\cite{maibier2016structure,defouw1989mapping,shumko1988vascular,nikiforidis1999quantitative,smith2016microvascular}. The \ac{CAM} model also has been extensively employed for \textit{in vivo} experiments to investigate angiogenesis~\cite{staton2004current,storgard2005angiogenesis}, and human tumor growth and therapy~\cite{ribatti2008chick,mesas2024experimental}. Human cells, PDX, and tissue can be engrafted onto the \ac{CAM}, e.g., to study the effect of potential therapeutics, or to investigate angiogenic and anti-angiogenic processes involving tissue and cells. As a result, numerous substances have been reported to either promote or inhibit angiogenesis in the CAM model, including growth factors, anti-cancer agents, pro-angiogenic molecules, natural and synthetic compounds, antibodies, organometallic compounds, and antibiotics~\cite{chase2017development}.

\subsection{Benefits and Limitations}
In general, the CAM model is considered to be a flexible pharmaceutical testing platform with several advantageous and disadvantageous features~\cite{chu2022applications,chen2021utilisation}.
The CAM model provides the following benefits. First, the chick embryo grows into a fully developed organism, with vital organs such as heart and liver. The \ac{CAM} model provides nutrients and has impressive capabilities for angiogenesis, making it an ideal tool for modeling more complex biological systems. During early development, chick embryos lack a fully developed immune system, which not only reduces costs compared to using immune-compromised animals but also opens up new avenues for experimentation. Furthermore, CAM experiments facilitate high-throughput screening~\cite{zhou2013novel,dunker2019implementation}, allowing researchers to visualize results in real time. In addition, each egg can be used for multiple tests, and the highly vascularized nature of the \ac{CAM} creates a perfect backdrop for studying tumor growth. Moreover, CAM research is cost-effective, as the eggs are small (6~\si{\centi\meter} length and 4~\si{\centi\meter} diameter), and easy to replace. Finally and maybe most importantly, the \ac{CAM} model fulfills the 3R principles for the reduction, refinement, and replacement of animal models for research purposes~\cite{palumbo2023cam}. 

Despite these advantages of the CAM model, there are also some drawbacks, of course. One significant challenge is the rapid growth of the embryo, which necessitates continuous monitoring to capture developmental changes. However, continuous monitoring is challenging, as the chick embryo cannot be kept outside the incubator for more than a few hours. Additionally, distinguishing between the established vascular network and newly formed capillaries can be challenging. Moreover, the CAM is quite sensitive to its surroundings; even slight changes in pH, temperature, or oxygen levels can impact experimental results. A further drawback can occur during drug studies in the \ac{CAM} model, as the way drugs are metabolized in the CAM can differ from how they are processed by mammals~\cite{nowak2014chicken}. 

\subsection{Main Research Directions}
In the \ac{CAM} literature, two main research directions are widely recognized: (i) preclinical studies -- with a particular emphasis on angiogenesis, cancer research, and drug screening -- and (ii) the development of imaging and quantification techniques for vascular analysis and drug monitoring.

Due to its strong vascularization, the \ac{CAM} model has become a valuable platform for drug screening~\cite{pawlikowska2020exploitation,lange2001new} and vascular-targeting therapies~\cite{li2011dual}. It enables rapid evaluation of compounds affecting angiogenesis and vascular integrity. Recent advances have established the CAM model as a robust system for PDX implantation, supporting the long-term observation of tumor vascularization. This application is particularly impactful in cancer research, providing insights into tumor biology and facilitating biomarker studies~\cite{Wagner24, eichhorst2024establishment}.

The research on the \ac{CAM} model has also driven the development of imaging methods for the analysis of vascular growth and drug effects~\cite{schueler2024ultra}. These advancements allow for the quantification of vessel density, structure, and blood flow, supporting angiogenesis studies and drug efficiency assessments~\cite{mcdonald2003imaging}. While traditional destructive methods like histology and stereology remain common, non-destructive approaches -- including X-ray, MRI, ultrasound, optical imaging, and AI-assisted analysis -- offer detailed visualization of vascular networks~\cite{geng2018hatching, verhoelst2011structural}. These tools enable both structural and functional assessments, such as measuring blood flow and oxygenation, especially in tumor-bearing CAM models for evaluating vascular-targeted therapies.

\vspace*{-0.2cm}
\subsection{Topology}
\label{sec:topol}
As shown in Fig.~\ref{fig:cam-schematic}, the \ac{CAM} model includes three main elements: the chick embryo, the yolk bag, and the \ac{CAM}. 
As previously described, the \ac{CAM} is an extraembryonic membrane that is formed by the mesodermal fusion of allantois and chorion within the third and tenth \ac{DED} \cite{kundekova2021chorioallantoic, nowak2014chicken}. The \ac{CAM} is comprised of a vast array of blood vessels and a dense capillary network. The vascular structure developing in the \ac{CAM} is completely connected to and supplied by the embryo’s vascular system and -- depending on the \ac{DED} -- also to the organs of the embryo. The yolk bag is needed to nourish the embryo and takes over the function of organs such as the liver, until these organs have fully developed in the embryo.
Therefore, the \ac{CAM} model represents a realistic \textit{in vivo} model of a circulatory system providing a highly complex branched closed-loop vascular system as well as several natural environmental effects such as various clearance mechanisms caused by organs.

\begin{figure*}[t!]
    \centering
    \begin{minipage}{0.19\linewidth}
        \includegraphics[width=\linewidth]{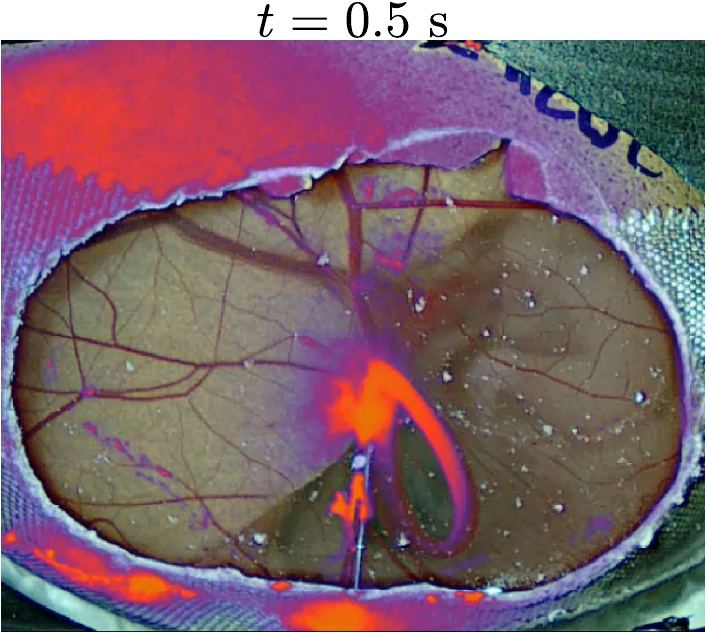}
    \end{minipage}\hfill
    \begin{minipage}{0.19\linewidth}
        \includegraphics[width=\linewidth]{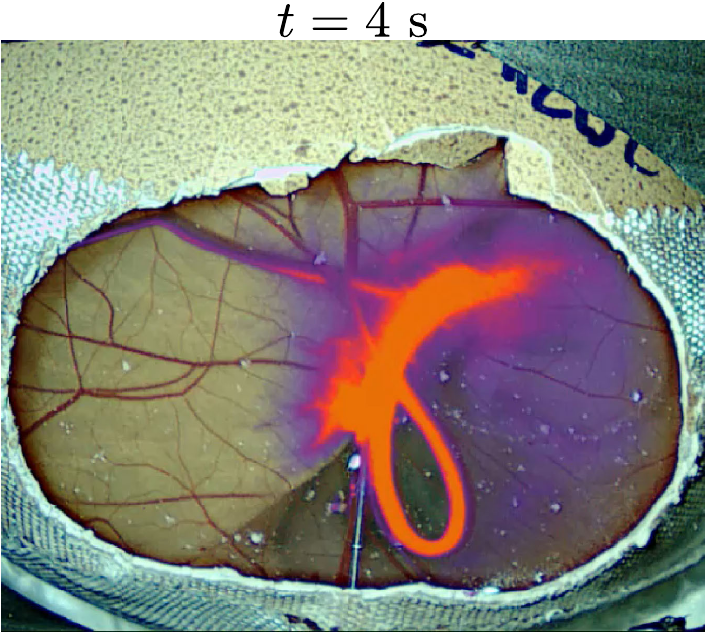}
    \end{minipage}\hfill
    \centering
    \begin{minipage}{0.19\linewidth}
        \includegraphics[width=\linewidth]{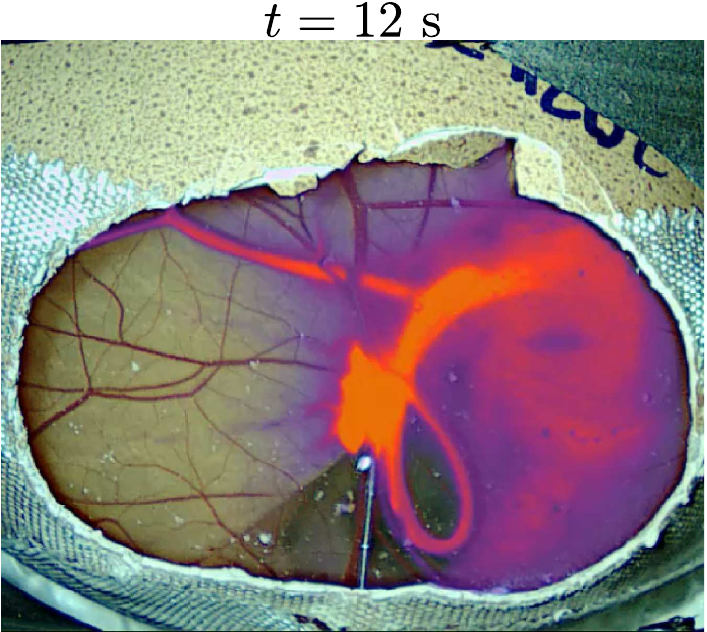}
    \end{minipage}\hfill
    \begin{minipage}{0.19\linewidth}
        \includegraphics[width=\linewidth]{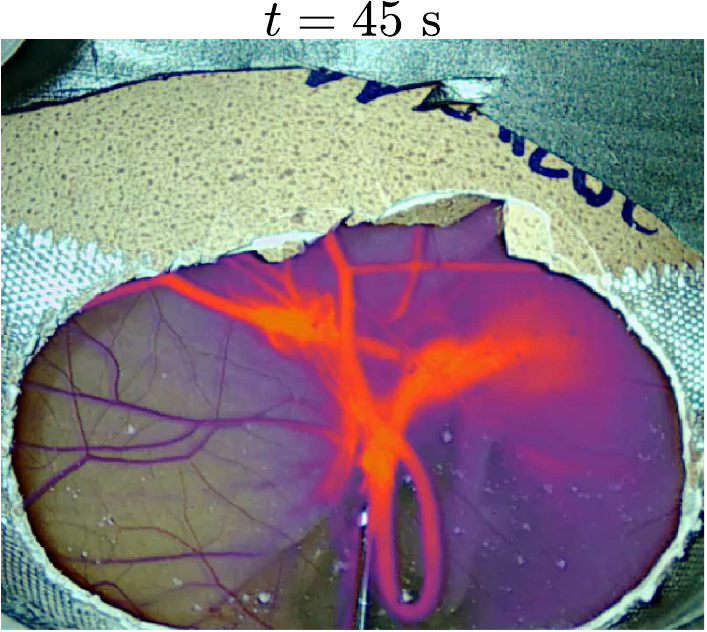}
    \end{minipage}\hfill
    \centering
    \begin{minipage}{0.19\linewidth}
        \includegraphics[width=\linewidth]{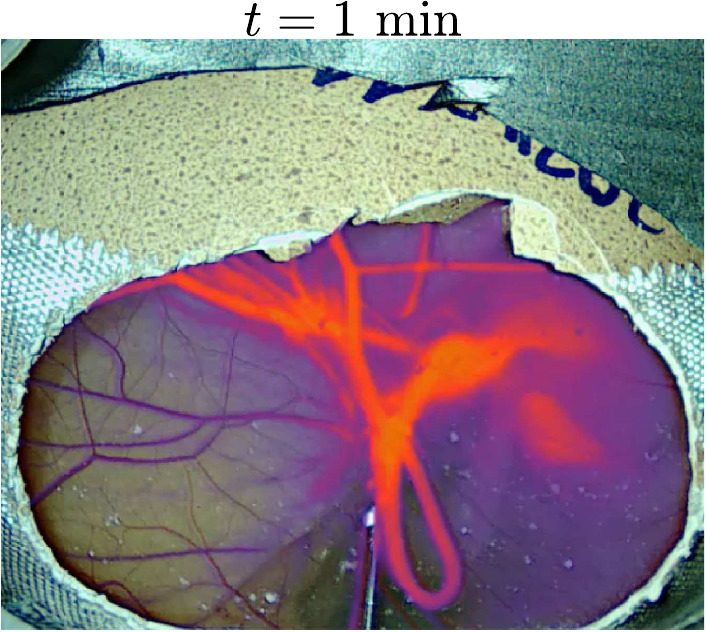}
    \end{minipage}\\
    \vspace{0.1cm}
    \centering
    \begin{minipage}{0.19\linewidth}
        \includegraphics[width=\linewidth]{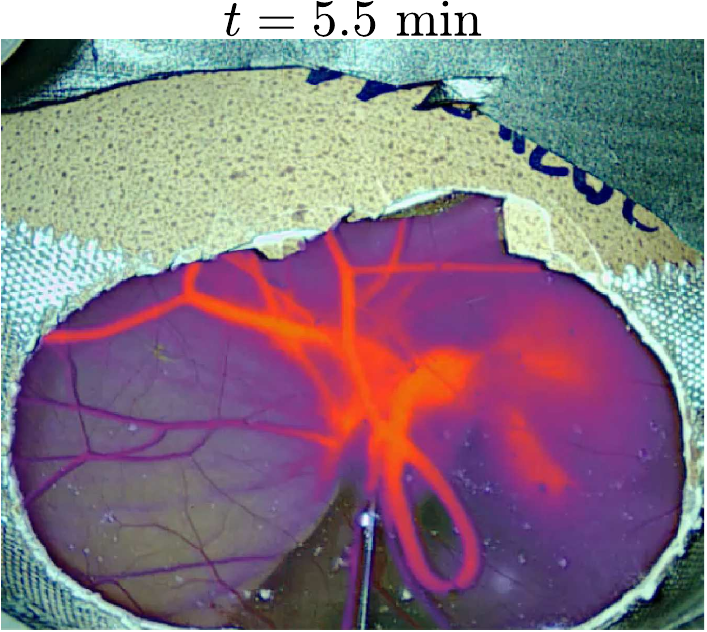}
    \end{minipage}\hfill
    \begin{minipage}{0.19\linewidth}
        \includegraphics[width=\linewidth]{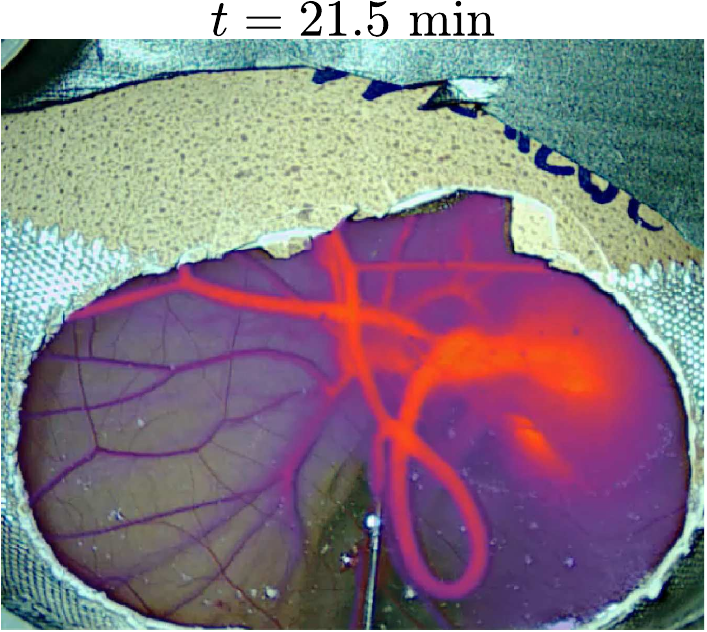}
    \end{minipage}\hfill
    \centering
    \begin{minipage}{0.19\linewidth}
        \includegraphics[width=\linewidth]{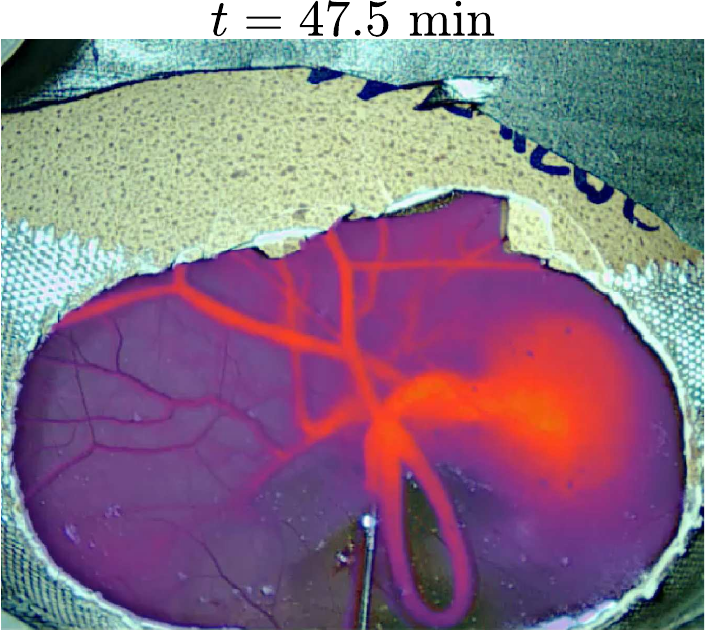}
    \end{minipage}\hfill
    \begin{minipage}{0.19\linewidth}
        \includegraphics[width=\linewidth]{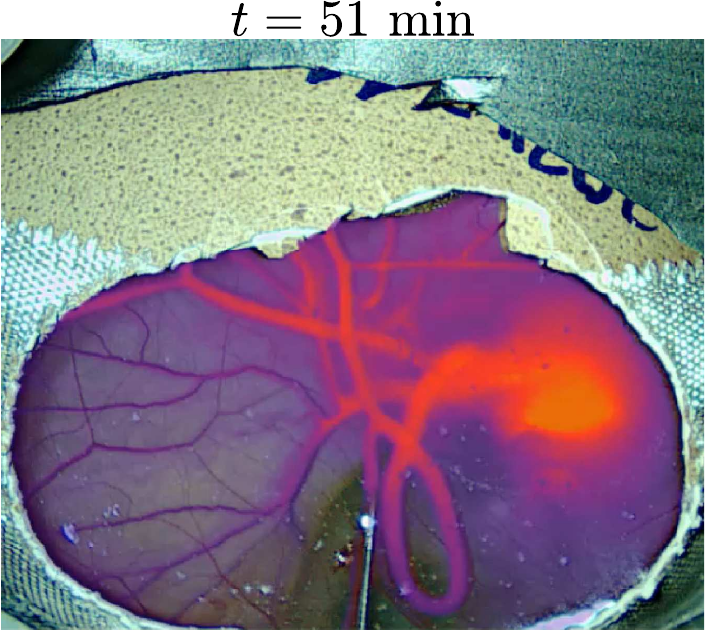}
    \end{minipage}\hfill
    \centering
    \begin{minipage}{0.19\linewidth}
        \includegraphics[width=\linewidth]{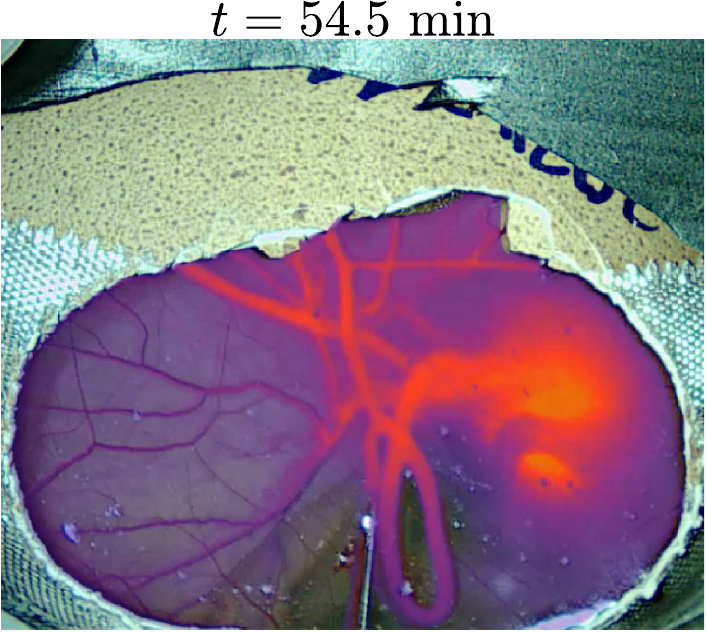}
    \end{minipage}    
    \caption{\small Photos of the \ac{ICG} fluorescence intensity distribution and accumulation in the \ac{CAM} model over time for a long-term experiment.}
    \label{fig:Liver}
\end{figure*}

\vspace*{-0.2cm}
\subsection{Particle Propagation}
\label{sec:sec2B}
Characterizing the influence of diffusion, flow, and vascularization on the propagation of particles inside the \ac{CAM} model is not straightforward. One of the biggest challenges is the fact that the \ac{CAM} is a constantly developing system \cite{Makanya2016, Guerra2021}. Depending on the \ac{DED}, the vascular system in the \ac{CAM} and in the embryo is changing significantly and therefore, the blood volume, average volume flux, and also other parameters depend on the \ac{DED} \cite{Tazawa1977}. Also, the effective local blood flow velocity in the \ac{CAM} model strongly depends on the \ac{DED} and possibly exhibits a large variance among different eggs. In \cite{Kloosterman2014}, it has been shown that the flow profile is laminar and measured time-averaged mean velocities range from $1~\si{\micro\meter\per\second} - 1~\si{\milli\meter\per\second}$ for vessel diameters ranging from $25 - 500~\si{\micro\meter}$. In \cite{Maibier2016}, it has been additionally shown that the topology of the \ac{CAM} can be divided into feeding arterioles and draining venules having larger diameters, and secondary arterioles and venules with normally distributed diameters having means of $46.3~\si{\micro\meter}$ and $49~\si{\micro\meter}$, respectively. Moreover, the authors of \cite{Maibier2016} showed that the measured effective blood flow velocities increase linearly with the vessel diameter \cite[Fig.~4]{Maibier2016}. The influence of diffusion inside the \ac{CAM} model depends on the properties of the particles injected and the occurring chemical reactions, e.g., the affinity and binding of molecules to serum proteins such as albumin \cite{Weixler23}.

\begin{figure}[t]
    \centering
    \includegraphics[width=0.8\linewidth]{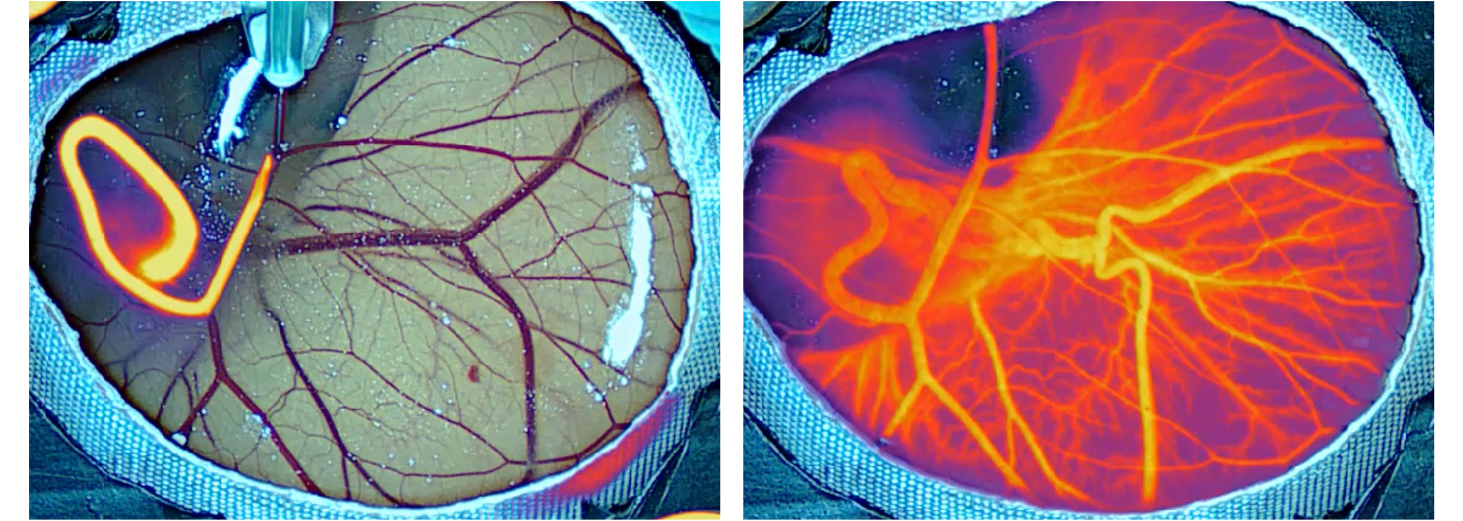}
    \caption{\small Snapshots of the distribution of fluorescent molecules in the vascular system of the \ac{CAM} model directly after injection via a syringe (left) and after a few seconds fully distributed (right). 
    }
    \label{fig:icg-injection}
\end{figure}

Depending on the \ac{DED}, the yolk bag and the organs of the chick embryo can have a strong influence on the propagation of particles in the \ac{CAM} model, as they introduce additional clearance and accumulation effects. As in the first \acp{DED}, the yolk bag takes over the functions of the liver to filter toxic particles, particles entering the yolk bag will accumulate with a certain probability. In later \acp{DED}, organs in the embryo (e.g., brain and liver) start to develop and particles accumulate there, too \cite{Ettner2024, xiao:biol:2024}. \ac{ICG} molecules accumulate preferably in the liver of the embryo as hepatocytes actively take up the particles through specialized transport mechanisms, and they resist being broken down by the metabolic processes~\cite{yaseen2008vivo, Givisiez2020}. Inside hepatocytes, \ac{ICG} is quickly released into bile without being altered, and it does not return to the bloodstream or get recycled through the liver and intestines. 
%
Figure~\ref{fig:Liver} shows the \ac{ICG} fluorescence intensity at different time instances after particle injection to further illustrate the particle propagation and accumulation processes inside the \ac{CAM} model. After injection, the particles quickly distribute inside the vascular system. Then, the fluorescence intensity starts to increase at the location of the embryo (right side of the egg) for $t > 5.5~\si{\minute}$. At the same time, the \ac{ICG} fluorescence intensity in the vascular system decreases. In particular, the \ac{ICG} particles gradually move from the vessels into the organs of the embryo where they accumulate, i.e., considering the age of the embryo shown in Fig.~\ref{fig:Liver} (\ac{DED} $> 10$), particles preferably accumulate in the liver of the embryo \cite{xiao:biol:2024, Givisiez2020}. 

\subsection{Tumor and Distribution Studies} 
\label{subsec:cam-tumor}
The \ac{CAM} model has become increasingly important for tumor research over the few last years. As previously described, it is employed to gain more insights into tumor biology, metastasis, drug distribution, and molecular screening \cite{Valdes2002, pion20223d, nowak2014chicken}. Besides for research on tumor growth and metastasis \cite{Miebach22}, the CAM model is also utilized for analyzing the distribution of specific molecules during the development and testing of targeted therapeutics \cite{Ettner2024, pawlikowska2020exploitation}.

\begin{figure}[t]
    \centering
    \includegraphics[width=0.8\linewidth]{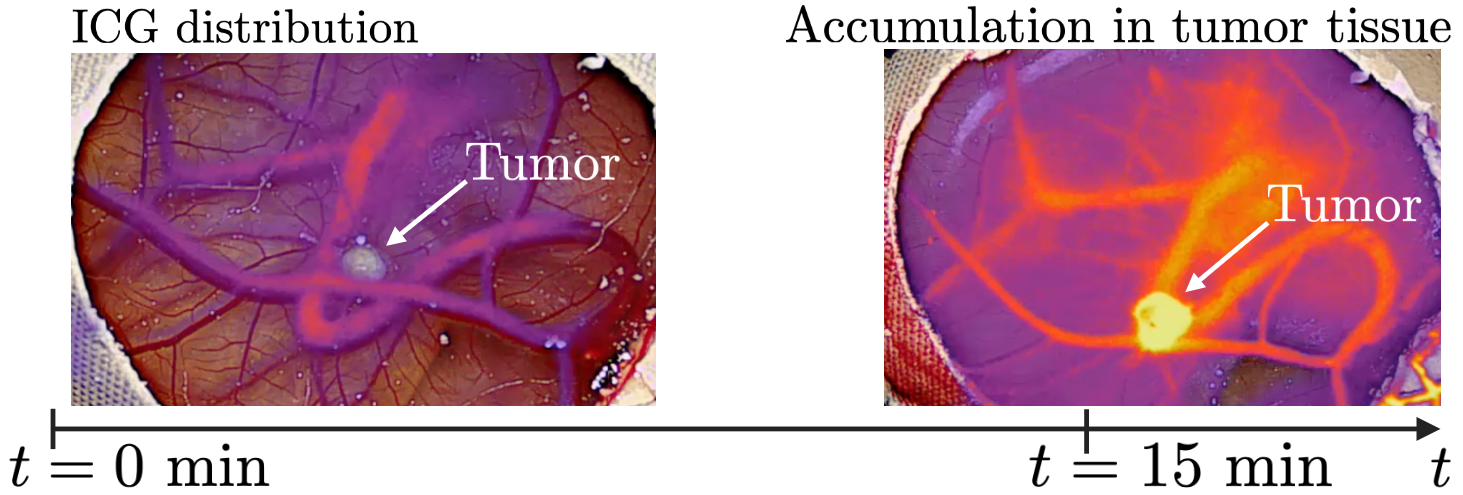}
    \caption{\small \ac{CAM} model for tumor and distribution studies. Left: \ac{ICG} injection into \ac{CAM} vessels at $t = 0~\si{\minute}$ and distribution in the vascular system. Right: \ac{ICG} is accumulated in tumor tissue after $\sim \!15~\si{\minute}$ \cite{Ettner2024}.}
    \label{fig:cam-drug-study}
\end{figure}

One key challenge is understanding how molecules distribute and accumulate, as well as visualizing tumors effectively. To tackle this, fluorescence imaging offers a solution. It relies on injecting fluorescent molecules into the \ac{CAM} model \cite{Zucal21, Zucal22,Ettner2024}. A fluorescent molecule, commonly used in the \ac{CAM} model, is the near-infrared fluorescent agent \ac{ICG}, which is mainly used for cancer imaging and image-guided surgeries in clinical settings \cite{Weixler23}. Stimulated with light in the near-infrared range ($750~\si{\nano\meter}$ and $950~\si{\nano\meter}$), \ac{ICG} starts to fluoresce, allowing visualization of vascular and anatomical structures (see Fig.~\ref{fig:icg-injection}). Due to its chemical composition, \ac{ICG} interacts with membrane and macro-molecular serum proteins, including albumin, which is commonly enriched in many cancers \cite{Weixler23, Ettner2024}. 
Therefore, \ac{ICG} is often used in clinical practice for the intraoperative visualization of tumors \cite{Chauhan23}. Moreover, as \ac{ICG} preferably accumulates in tumor tissue and partly shows similar properties as drug molecules, it can be employed in the \ac{CAM} model to characterize and optimize the accumulation of molecules in tumor tissue as a basis for the design and optimization of drug delivery systems \cite{Weixler23,Chauhan23,Ettner2024}. Figure~\ref{fig:cam-drug-study} shows an example of the \ac{ICG} distribution in the vascular system of the \ac{CAM} model and its accumulation in tumor tissue \cite{Ettner2024}. This accumulation is indicative of the tumor's enhanced vascular permeability, which is a characteristic feature of the growing tumor microenvironment. The ability to visualize \ac{ICG} uptake in real time provides valuable insight into the tumor's vascular structure and can be useful for assessing therapeutic interventions targeting tumor.

\section{Channel Modeling}
\label{sec:particleDis}
In this section, we derive an \ac{MC} channel model for the particle propagation in dispersive closed-loop systems which is subsequently used to approximate the particle distribution inside the \ac{CAM} model. To this end, first, we  provide an interpretation of the \ac{CAM} model as a closed-loop \ac{MC} system. Then, based on observations from  experimental measurements of the distribution and accumulation of particles inside the \ac{CAM} model, we present our modeling assumptions. These assumptions are used to derive an analytical model for the molecule distribution in closed-loop systems. Finally, we validate the analytical model by a comparison to results obtained by~\ac{PBS}.

\begin{figure}[t]
    \centering
    \includegraphics[width=0.6\linewidth]{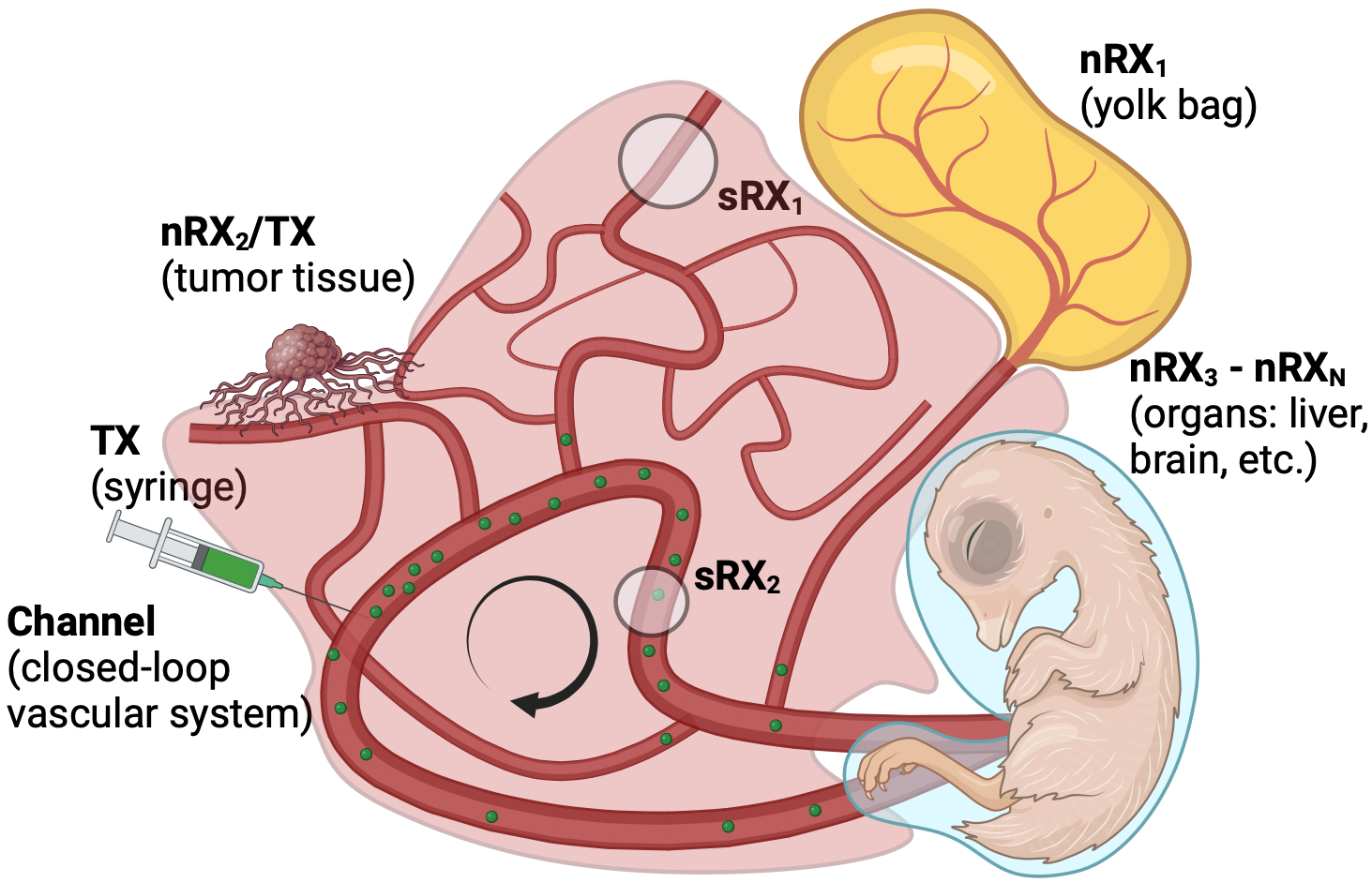}
    \caption{\small Interpretation of the \ac{CAM} model as closed-loop \ac{MC} system with \acp{TX}, e.g., a syringe or a tumor, several \acp{nRX}, e.g., yolk bag, organs, tumor tissue, and several \acp{sRX} (Created with \url{BioRender.com}).}
    \label{fig:cam-mc}
\end{figure}

\subsection{CAM Model as an MC System}
\label{sec:cam-mc}
Figure~\ref{fig:cam-mc} shows a representation of the \ac{CAM} model from Fig.~\ref{fig:cam-schematic} rearranged as an \ac{MC} system. The system consists of \acp{TX}, e.g., a syringe or a tumor, and possibly multiple \acp{RX} where we distinguish between \acfp{nRX}, such as the yolk bag, organs, and tumor tissue, and \acfp{sRX} such as localized measurement units. Between the \acp{TX} and \acp{RX}, the closed-loop vascular system of the \ac{CAM} represents the channel. 
\begin{figure}[t!]
    \centering
    \begin{minipage}{0.49\linewidth}
        \includegraphics[width=\linewidth]{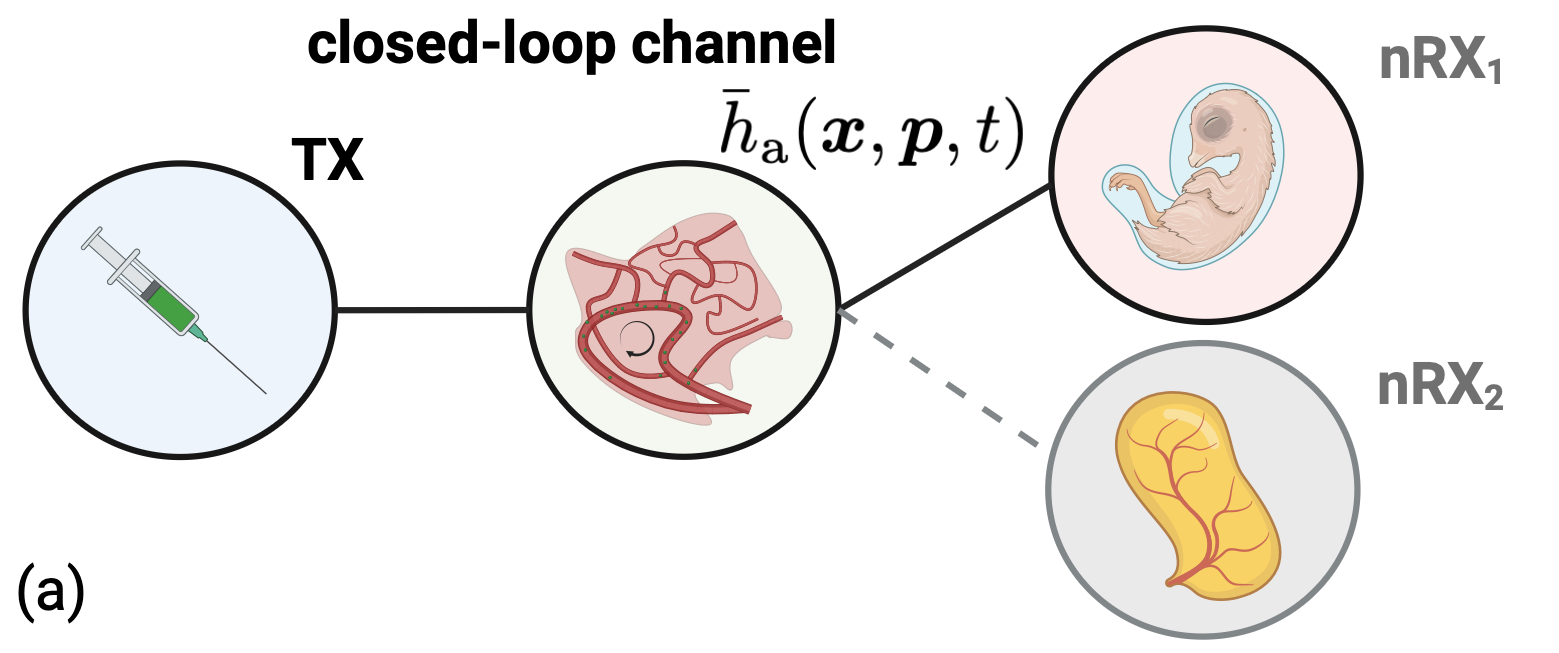}
    \end{minipage}\hfill
    \begin{minipage}{0.49\linewidth}
        \includegraphics[width=\linewidth]{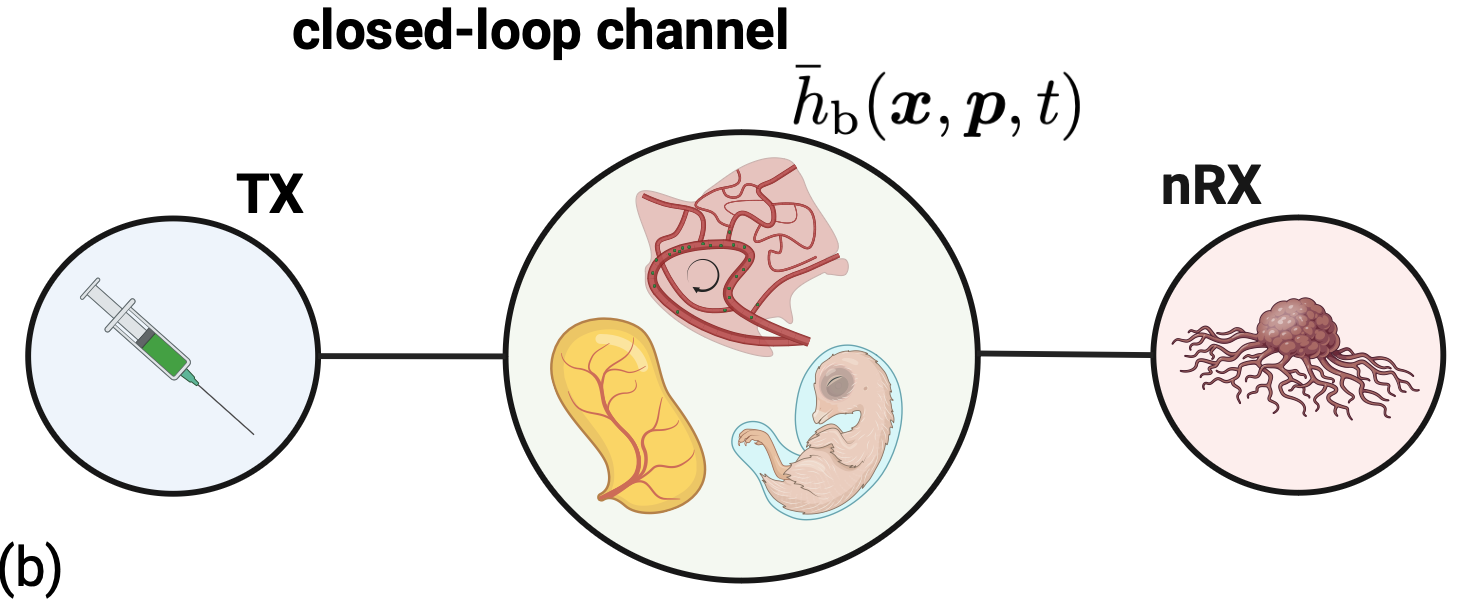}
    \end{minipage}\\
    \centering
    \includegraphics[width=0.5\linewidth]{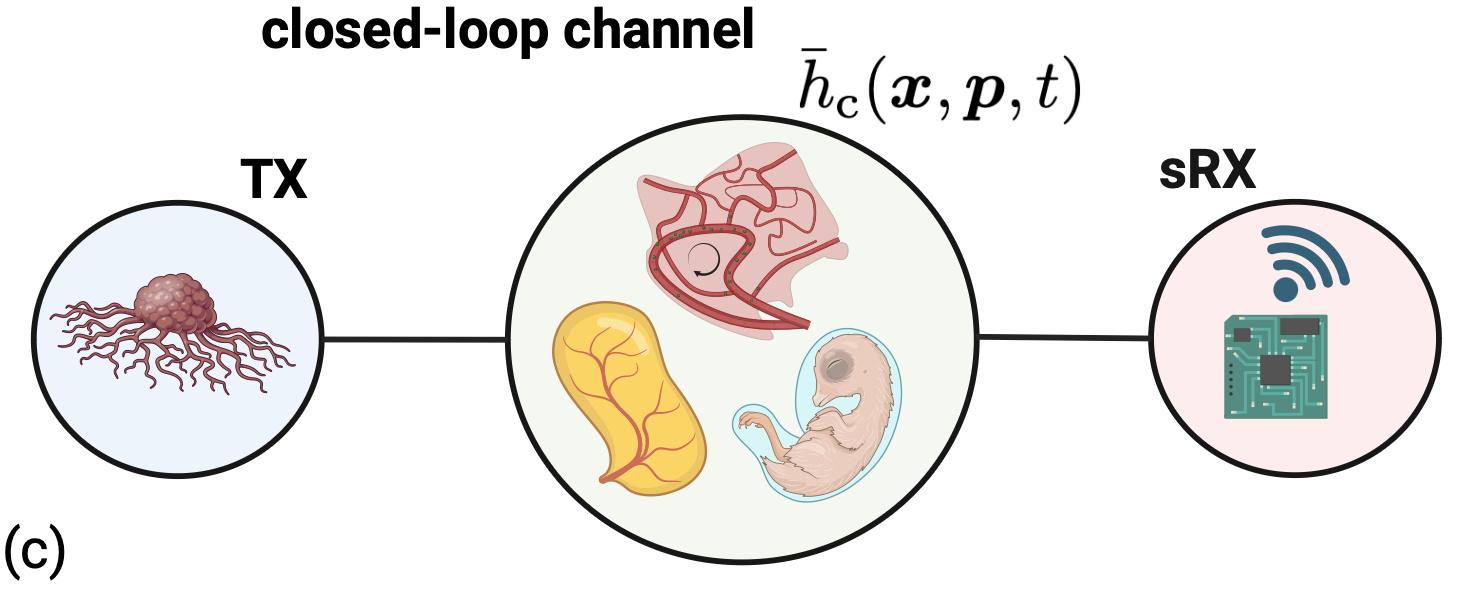}
    \caption{\small Application-dependent specializations of system in Fig.~\ref{fig:cam-mc}. (a): Study of molecule distribution and accumulation, yolk bag and organs are \acp{RX}. (b): Study of molecule accumulation in tumors, engrafted tumor tissue is the \ac{RX}, and the yolk bag and organs are part of the channel. (c): Anomaly detection or monitoring tasks, engrafted tumor tissue is the \ac{TX} (Created with \url{BioRender.com}).}
    \label{fig:cam-mc-spec}
\end{figure}

Depending on the targeted application, the topology in Fig.~\ref{fig:cam-mc} can be specialized as it is shown for three specific cases in Fig.~\ref{fig:cam-mc-spec}. Figure~\ref{fig:cam-mc-spec}~(a) shows the \ac{MC} system topology if the distribution and accumulation of molecules injected into the \ac{CAM} model are studied. In this scenario, depending on the \ac{DED}, the yolk bag and organs can be interpreted as multiple \acp{RX}, where molecules accumulate. The closed-loop vascular system is the channel, which can be described by a time-variant, possibly nonlinear function $\bar{h}_\mathrm{a}(\bm{x},\bm{p},t)$, depending on a three-dimensional space variable $\bm{x}$, a set of environmental parameters $\bm{p}$ (e.g., diffusion coefficient, velocity, length of the vessel), and time $t$. Hereby, the time-variance of the channel mainly arises from random movements of the embryo, its growth, and the non-constant blood flow.
Figures~\ref{fig:cam-mc-spec}~(b) and (c) show ``point-to-point'' \ac{MC} systems, where a \ac{TX} releases signaling molecules for an \ac{nRX} or \ac{sRX}. Figure~\ref{fig:cam-mc-spec}~(b) shows a drug delivery system as a particular application, i.e., the \ac{TX} injects drug molecules into the \ac{CAM} model and the accumulation in the tumor tissue can be first characterized and later optimized \cite{Ettner2024}. In this scenario, the \ac{nRX} is human tumor tissue engrafted onto the \ac{CAM}, see Fig.~\ref{fig:cam-drug-study}, and the organs and yolk bag are part of the closed-loop channel $\bar{h}_\mathrm{b}(\bm{x},\bm{p},t)$ influencing the propagation of drug molecules by degradation effects. Figure~\ref{fig:cam-mc-spec}~(c) shows a health monitoring application, where tumor tissue, modulated by the \ac{TX}, releases specific biomarkers into the closed-loop channel $\bar{h}_\mathrm{c}(\bm{x},\bm{p},t)$, which can be detected by an \ac{sRX}. 
\subsection{Modeling Assumptions}
\label{subsec:assumpt}

The topology of the \ac{CAM} model is highly complex due to its vascularization and closed-loop character, and it is constantly changing over time and partly unknown. Therefore, the development of an exact model for the propagation of molecules is infeasible. So far, only individual vessel trees of the \ac{CAM} vascular system have been modeled explicitly and simulated numerically for hemodynamic studies \cite{Maibier2016, Kloosterman2014}. However, besides their high computational cost, such numerical models provide only limited insight into the overall dynamics and limited applicability for analytical characterization. Therefore, we focus on the development of insightful approximate models to characterize the molecule distribution inside the \ac{CAM} model. 

\begin{figure}[t!]
    \centering
    \includegraphics[width=0.99\linewidth]{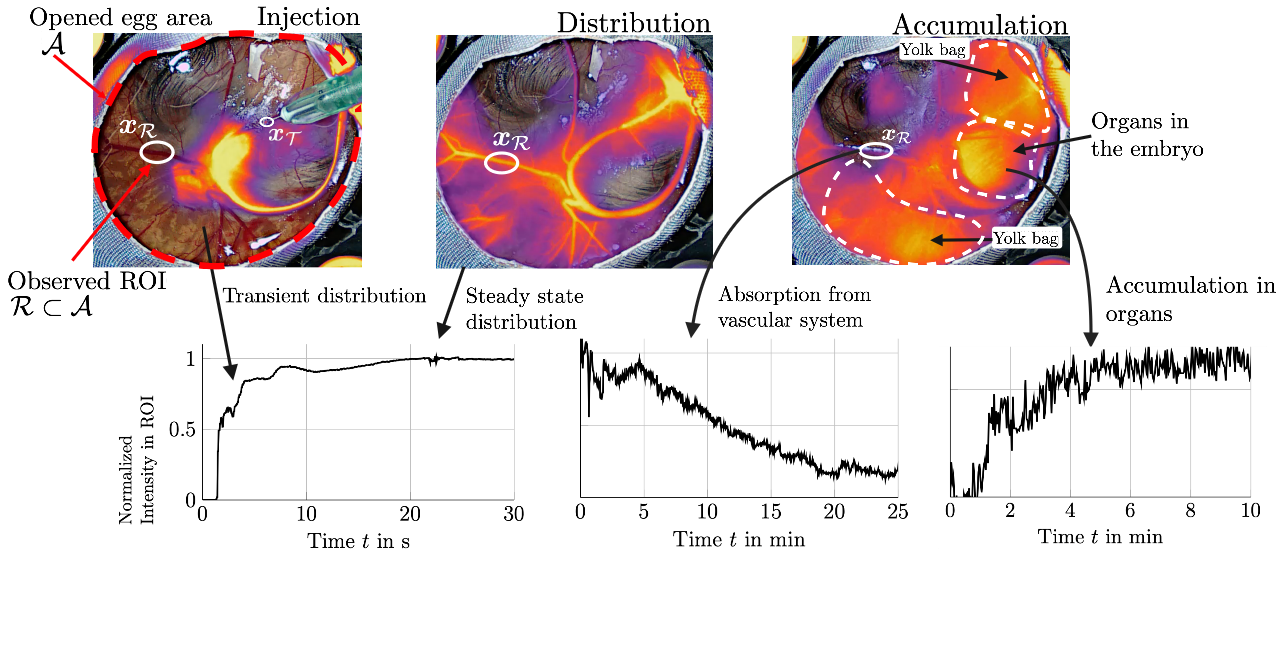}
    \caption{\small Top (from left to right): Injection and distribution of \ac{ICG} molecules in the vascular system and accumulation in the yolk bag and organs. Bottom: Measured fluorescence intensity in different ROIs, such as vascular system and organs of the embryo (Created with Biorender.com).}
    \label{fig:dist-dec-overview}
\end{figure}
Figure~\ref{fig:dist-dec-overview} shows experimental results for the distribution of the fluorescent molecule \ac{ICG} inside the \ac{CAM} model at different time instances (top of Fig.~\ref{fig:dist-dec-overview}) and the fluorescence measured over time in \acf{ROI} $\varcal{R}$ centered at $\bm{x}_\varcal{R} \in \mathcal{A}$, where $\varcal{A}$ is the opened egg area observed by an \ac{ICG} fluorescence camera. The setup corresponds to an \ac{MC} system where the tip of the syringe injecting the \ac{ICG} is the \ac{TX} located at $\bm{x}_\varcal{T}$, and the observation of fluorescence intensity $I(\bm{x}_\varcal{R},t)$ in the \ac{ROI} corresponds to a transparent \ac{sRX} located at $\bm{x}_\varcal{R}$ (bottom of Fig.~\ref{fig:dist-dec-overview}). Assuming that the intensity of the \ac{ICG} fluorescence is proportional to the \ac{ICG} molecule concentration~\cite{Chon2023}, from the figure, we make two main observations: 
\begin{itemize}
    \item[(i)] From the bottom left plot in Fig.~\ref{fig:dist-dec-overview} we observe that, after the injection, the \ac{ICG} molecule distribution reaches a steady state inside \ac{ROI} $\varcal{R}$ very quickly (after approximately $25~\si{\second}$). This is due to the laminar flow, dispersion, and turbulences introduced by the high vascularization and the closed-loop character of the \ac{CAM}'s vascular system \cite{Kloosterman2014, Maibier2016}.
    \item[(ii)] From the bottom middle plot in Fig.~\ref{fig:dist-dec-overview} we observe that, after the steady state has been reached, the \ac{ICG} molecules are absorbed from the vascular system into the yolk bag and organs of the \ac{CAM} model, leading to fewer molecules in the \ac{ROI} $\varcal{R}$. Compared to the initial molecule distribution process, this molecule absorption process is much slower and occurs on the order of minutes. The bottom right plot in Fig.~\ref{fig:dist-dec-overview} shows the accumulation of the \ac{ICG} molecules in the embryo's organs. Concurrently, we can observe that the fluorescence intensity increases in the organ's region as it decreases in the vascular system.
\end{itemize}
Based on observations (i) and (ii), we divide the temporal dynamics of the particle distribution in the \ac{CAM} model into three phases, i.e., the \textit{transient phase}, \textit{steady state phase}, and \textit{accumulation phase}. The \textit{transient phase} describes the fast initial spread of molecules inside the \ac{CAM} model's vascular system. In practice, this phase is also affected by the dynamics of the particle injection mechanism. The \textit{steady state phase} represents a temporary steady state, which occurs after the \textit{transient phase} once the injected particles are almost equally distributed in the vascular system. Finally, the \textit{accumulation phase} represents all accumulation phenomena of particles, e.g., in the yolk bag and organs, whose dynamics are much slower than the transient distribution process (see Fig.~\ref{fig:dist-dec-overview}).
Thus, we characterize the dynamics of the observed \ac{ICG} intensity $I(\bm{x}_\mathcal{R},t)$ at any point $\bm{x}_\mathcal{R}\in\mathcal{A}$ as follows
\begin{align}
    I(\bm{x}_\mathcal{R},t) =
    \begin{cases}
        I_{\mathrm{dist}}(\bm{x}_{\mathcal{R}},t) & t \leq t_\mathrm{acc}\Hquad,\\
        I_{\mathrm{acc}}(\bm{x}_{\mathcal{R}},t) & t > t_\mathrm{acc}\Hquad,
    \end{cases}
    \label{eq:phases}
\end{align}
where $I_{\mathrm{dist}}(\bm{x}_{\mathcal{R}},t)$ represents the \textit{transient} and \textit{steady state phases}, $I_{\mathrm{acc}}(\bm{x}_{\mathcal{R}},t)$ represents the \textit{accumulation phase}, and $t_\mathrm{acc}$ is the time where the \textit{steady state phase} ends and the \textit{accumulation phase} starts. Based on the division in \eqref{eq:phases}, we develop separate parametric models for the \textit{transient} and \textit{steady state phases} on the one hand and for the \textit{accumulation phase} on the other hand. In the following, we derive a physically motivated analytical model for the particle distribution in a closed-loop system, which will be used as part of the parametric model to approximate $I_{\mathrm{dist}}(\bm{x}_{\mathcal{R}},t)$ in the \textit{transient} and \textit{steady state phases}. 
Since the model for the \textit{accumulation phase} proposed in this paper is not ``physically motivated'', and we assume it is independent from the \textit{transient} and \textit{steady state phases}, the parametric model for the approximation of $I_{\mathrm{acc}}(\bm{x}_{\mathcal{R}},t)$ during the \textit{accumulation phase} is presented in Section~\ref{sec:est}.
\begin{figure}[t]
    \centering
    \includegraphics[width=0.8\linewidth]{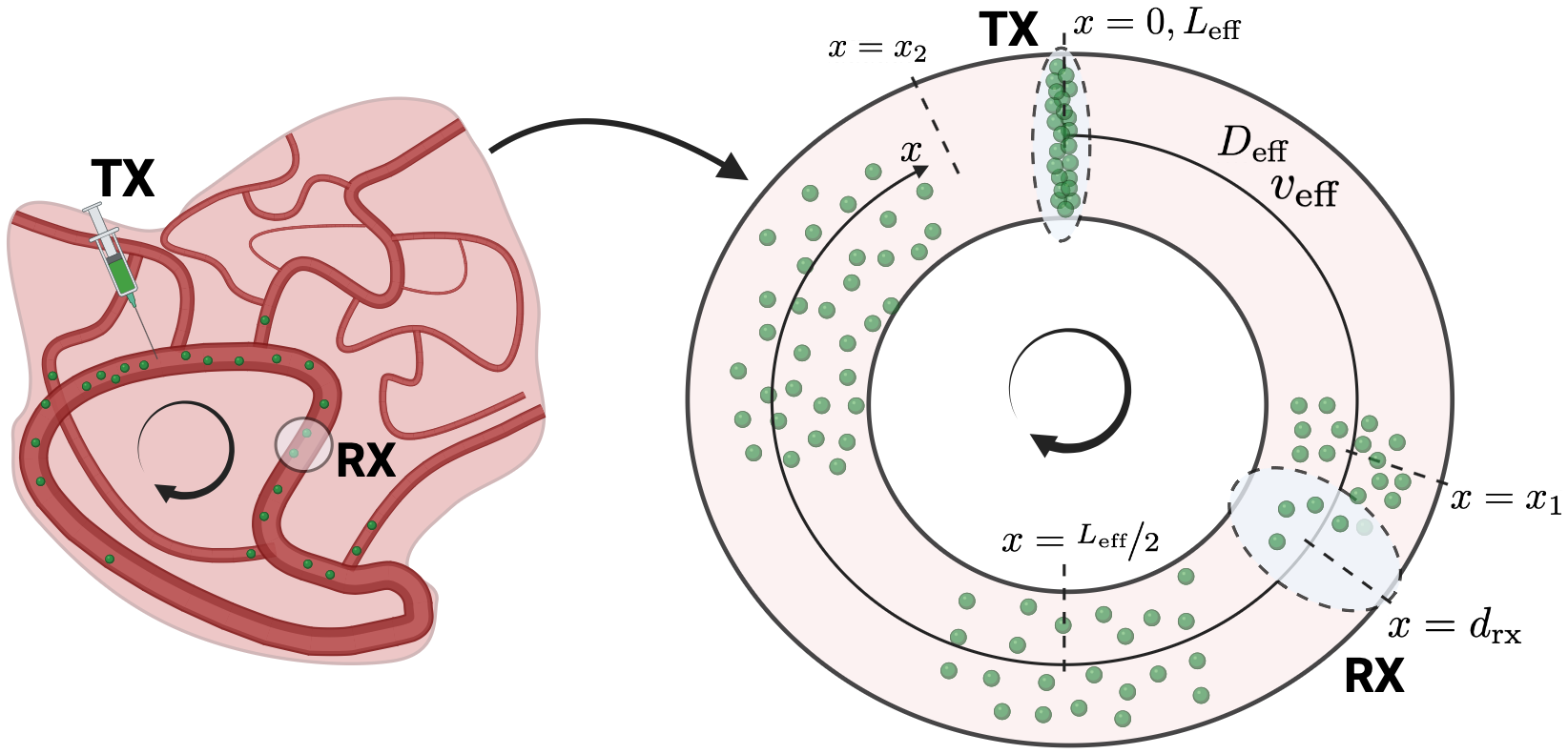}
    \caption{\small Proposed approximative model for the distribution of molecules in closed-loop systems. Left: Closed-loop \ac{CAM} vascular system. Right: Approximation as a closed-loop pipe with effective diffusion coefficient $\Deff$ and effective velocity $\veff$ (Created with \url{BioRender.com}).}
    \label{fig:cam-mc-approx}
\end{figure}

\subsection{Diffusion and Flow in Closed-loop Systems}\label{sec:3B}
In order to develop a model for the \textit{transient} and \textit{steady state phases}, we first propose an analytical model for the propagation of molecules in dispersive closed-loop systems, which we will use in Section~\ref{sec:exp} as a parametric model to approximate the propagation of molecules between a \ac{TX} and an \ac{RX} inside the \ac{CAM}'s vascular network. 
As previously described, the exact modeling of the topology of the \ac{CAM}'s vascular system is infeasible. Therefore, as shown in Fig.~\ref{fig:cam-mc-approx}, we approximate the $3$D closed-loop highly vascularized system by a $3$D closed-loop pipe of length $\Leff$. In the closed-loop pipe, molecules are transported by diffusion and flow. Thus, due to the high vascularization, we expect that the molecule transport inside the \ac{CAM}'s vascular system is highly dispersive. Therefore, we assume that molecule propagation occurs in the dispersive regime \cite{Jamali2019}, and model the 3D pipe on the right hand side of Fig.~\ref{fig:cam-mc-approx} by a 1D Aris-Taylor dispersion model. In this case, we can describe the concentration of molecules $p(x,t)$ in the closed-loop pipe system in Fig.~\ref{fig:cam-mc-approx} by a $1$D diffusion advection process as follows \cite{Probstein2005}
\begin{align}
    \partial_t p(x,t) = \Deff \partial_{xx} p(x,t) - \veff \partial_x p(x,t)\Hquad, 
    \label{eq:aris}
\end{align}
where $\partial_t$ and $\partial_{xx}$ denote first order partial time- and second order space derivative, respectively, and the space coordinate $x\in[0, \Leff]$ is along the pipe (see Fig.~\ref{fig:cam-mc-approx}). Parameters $\Deff$ in $\si{\square\meter\per\second}$ and $\veff$ in $\si{\meter\per\second}$ denote the effective diffusion coefficient and effective flow velocity, respectively.


Equation~\eqref{eq:aris} can be solved straightforwardly if we assume $\Leff\to\infty$ and a uniform molecule release at $x = 0$ and $t = 0$, yielding a normal distribution in $x$ \cite{Jamali2019}
\begin{align}
    p_\mathrm{n}(x,t) = \frac{1}{\sigma(t)\sqrt{2\pi}}\exp\left(-\frac{(x - \mu(t))^2}{2\sigma^2(t)}\right)\Hquad,
    \label{eq:arisInfinite}
\end{align}
which has variance $\sigma^2(t) = 2 \Deff t$ and is shifted along the $x$-axis by its mean $\mu(t) = \veff t$. Solution \eqref{eq:arisInfinite} is commonly applied for the modeling of dispersive \ac{MC} systems in infinitely long tubes \cite{Jamali2019,Wicke2018}. However, \eqref{eq:arisInfinite} cannot be applied for closed-loop systems with finite $\Leff$, such as the \ac{CAM} model and many other envisioned application environments of \ac{MC}, e.g., the human circulatory system \cite{Brand2024closed}. 

To extend \eqref{eq:arisInfinite} to a closed-loop system with finite $\Leff$ (see right hand side of Fig.~\ref{fig:cam-mc-approx}), \eqref{eq:aris} has to be restricted to the domain $[0, \, \Leff]$ with periodic boundary conditions, i.e., $p(0,t) = p(\Leff,t)$ and $\partial_x p(0,t) = \partial_x p(\Leff,t)$. Then, \eqref{eq:aris} characterizes the $1$D dispersive propagation of molecules on a circle with circumference $\Leff$. A solution to \eqref{eq:aris} with periodic boundary conditions for a uniform release at $x =0$ and $t = 0$ is the wrapped normal distribution, which can be obtained from wrapping \eqref{eq:arisInfinite} to a circular domain \cite{Mardia1999}  
\begin{align}
    p_\mathrm{wn}(x,t) = \frac{\sqrt{2\pi}}{\Leff\bar\sigma(t)}\sum_{k=-\infty}^{\infty}\exp\left(\frac{-(\bar x - \bar\mu(t) + 2\pi k)^2}{2\bar\sigma^2(t)} \right)\Hquad, 
    \label{eq:wrappedNormal}
\end{align}
where $\bar\sigma^2(t) = \lambda^2\sigma^2(t)$, $\bar\mu(t) = \lambda\mu(t)$, and $\bar x = \lambda x$ denote the variance, mean, and position mapped on a circle of circumference $\Leff$ with scaling parameter $\lambda = \frac{2\pi}{\Leff}$. 
The general behavior of \eqref{eq:wrappedNormal} is depicted in Fig.~\ref{fig:cam-mc-approx}. At time $t = 0$, molecules are concentrated at $x = 0$ and start to propagate, mainly driven by flow $\veff$. Due to diffusion, variance $\bar{\sigma}^2$ is increasing over time and molecules are getting more and more dispersed. For $t \to \infty$, variance $\bar{\sigma}^2(t) \to \infty$, and the molecules are uniformly distributed in the closed-loop system. In Section~\ref{sec:est}, we will use \eqref{eq:wrappedNormal} as an approximation for characterizing the channel characteristics $\bar{h}(\bm{x},\bm{p},t)$ of the \ac{CAM} vascular system, i.e., $\bar{h}(\bm{x},\bm{p},t) \approx p_\mathrm{wn}(x,t)$.

\begin{figure}
    \centering
    \includegraphics[width=0.9\linewidth]{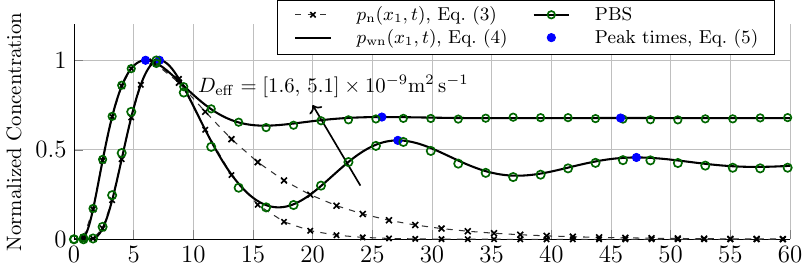}\\
    \includegraphics[width=0.9\linewidth]{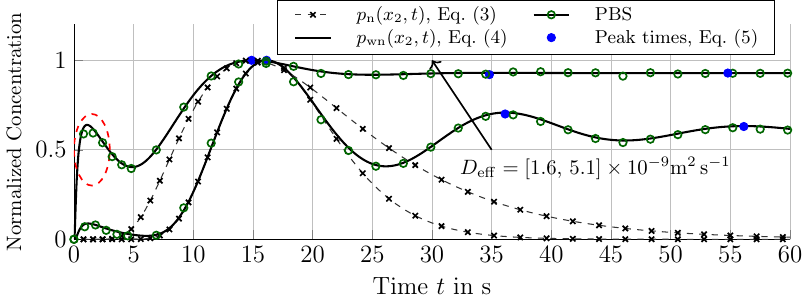}
    \caption{\small Comparison between $p_\mathrm{n}$ in \eqref{eq:arisInfinite} (for an infinite length pipe) and $p_\mathrm{wn}$ in \eqref{eq:wrappedNormal} (for a closed-loop pipe) at two different \ac{RX} locations $x_1$ (top) and $x_2$ (bottom) for different effective diffusion coefficients $\Deff$.}
    \label{fig:compDist}
\end{figure}

Inspecting $p_\mathrm{wn}(x,t)$ in \eqref{eq:wrappedNormal} and $p_\mathrm{n}(x,t)$ in \eqref{eq:arisInfinite}, it becomes clear that 
\eqref{eq:arisInfinite} corresponds to the term $k = 0$ on the right hand side of \eqref{eq:wrappedNormal} for $\lambda = 1$. The other terms with $k\neq 0$ in the sum in \eqref{eq:wrappedNormal} represent the contributions from multiple cycles in the closed-loop. For the scenario considered in Fig.~\ref{fig:cam-mc-approx}, the closed-loop also affects the received signal, as a single release may cause multiple peaks in the average number of molecules observed at the \ac{RX}. All peak times $\{t_\mathrm{max}(k, d_\mathrm{rx}) \vert k \in \mathbb{Z}\}$ observed by an \ac{RX} located at $d_\mathrm{rx}$ (see Fig.~\ref{fig:cam-mc-approx}) due to the molecules circulating multiple times in the loop, can be obtained by modifying the peak time for an infinitely long tube \cite{Wicke2018} as follows
\begin{align}
    t_\mathrm{max}(k, d_\mathrm{rx}) = \frac{\Deff}{\veff^2}\left(-1 + \sqrt{1 + \frac{\veff^2}{\Deff^2} (d_\mathrm{rx} + k\,\Leff)^2} \right)\Hquad.
    \label{eq:peakTime}
\end{align}

\subsection{Analysis and Validation}

To analyze the behavior of \eqref{eq:wrappedNormal}, we exemplarily consider a micro-scale closed-loop pipe system with length $\Leff=1~\si{\milli\meter}$, radius $r_0=100~\si{\micro\meter}$,  effective flow velocity $\veff=50~\si{\micro\meter\per\second}$, and two different diffusion coefficients $D = [1.25, \, 5]\times 10^{-9}~\si{\meter\per\second}$, yielding a system operating in the dispersive regime \cite{Schaefer2021}. The effective diffusion coefficient $\Deff$ can be calculated as $\Deff = D\big(1+\frac{1}{48}(\frac{r_{o}\veff }{D})^2\big)$~\cite[Eq.~(12)]{Wicke2018}. Molecules are released instantaneously at time $t = 0$ and position $x =0$. To validate \eqref{eq:wrappedNormal} and the modeling of the 3D pipe on the right hand side of Fig.~\ref{fig:cam-mc-approx} by a 1D drift-diffusion process, we compare our derived solution for the diffusion advection equation with periodic boundary condition to results from \ac{PBS}, shown as green circles in Fig.~\ref{fig:compDist}. 

We implemented the \ac{PBS} in a 3D straight cylinder with reflecting inner boundaries. The cylinder is elongated along the $x$ axis of a Cartesian coordinate system from $x=0$ to $x=\Leff$. We realize the closed-loop characteristic of the straight boundary by connecting the two circular bases of the straight cylinder, i.e., particles moving to $x>\Leff$ in a simulation step are moved to $x~\mathrm{mod}~\Leff$. At time $t = 0$, $10^4$ particles are uniformly distributed over a disc representing the boundary of the cylinder at $x=0$, and start diffusing for $t>0$. In addition, they are affected by the laminar flow in $x$ direction. The laminar flow we consider is radially symmetric with a parabolic profile, i.e., $ v_x = 2\veff\left( 1 - \left(r/r_0\right)^2\right)$,
where $v_x$, $r$, and $r_0$ are the particle's velocity in $x$ direction, the distance of the particle from the $x$ axis, and the radius of the cylinder. 
The PBS results (green circles) shown in Fig.~\ref{fig:compDist} were obtained by averaging over $20$ PBS realizations and a time step of $10^{-4}~\si{\second}$. 

Figure~\ref{fig:compDist} shows the normalized concentration obtained with~\eqref{eq:arisInfinite} and \eqref{eq:wrappedNormal} for two different \ac{RX} positions $x_1 = 0.39~\si{\milli\meter}$ and $x_2 = 0.84~\si{\milli\meter}$ (see also Fig.~\ref{fig:cam-mc-approx}). The approximate peak times according to \eqref{eq:peakTime} are indicated by blue dots. 
From Fig.~\ref{fig:compDist}, we observe that $p_\mathrm{wn}$ (black curves) successfully captures the effects introduced by a closed-loop system, such as the repeated observation of molecules by the \ac{RX} and the non-zero steady state concentration. In comparison, $p_\mathrm{n}$ (gray curve with markers) only captures the dynamics of the first peak as expected. An interesting effect, occurring in closed-loop systems is shown in the bottom plot of Fig.~\ref{fig:compDist} (highlighted by the red ellipse). For an \ac{RX} position closely behind the \ac{TX} and strong diffusion, an upstream peak occurs before the actual peak arrives at $t = 15~\si{\second}$. This peak originates from molecules diffusing against the direction of flow into the \ac{RX} region.
All results are in good agreement with the results from PBS, validating that \eqref{eq:wrappedNormal} is a suitable model for the propagation of molecules in dispersive closed-loop systems. 

\section{Parametric Models and Curve Fitting}
\label{sec:est}
In this section, we use the assumptions and the analytical model from the previous section as a foundation. We then develop parametric models for particle injection and distribution in the \ac{CAM} model during the \textit{transient}, \textit{steady state}, and \textit{accumulation phases}. Moreover, we describe the employed fitting and parameter estimation methods and metrics to evaluate the performance of the proposed parametric models. The proposed parametric models are fitted to experimental measurement data in Section~\ref{sec:exp} to show their capability to approximate the particle distribution in the \ac{CAM} model. 

\subsection{Parametric Models}\label{sec:ParamMod}
As previously mentioned, the \textit{transient phase} is strongly influenced by the particle injection dynamics. Therefore, we first present a parametric model to approximate the actual injection dynamics $f_\mathrm{inj}(t)$. Then, we propose two parametric models to approximate the particle concentration $I_\mathrm{dist}(\bm{x}_{\mathcal{R}},t)$ during the \textit{transient} and \textit{steady state phases}, based on the analytical model developed in Section~\ref{sec:3B}. Finally, we propose a first simple model to approximate the particle accumulation $I_\mathrm{acc}(\bm{x}_{\mathcal{R}},t)$ in the organs of the embryo within the \ac{CAM} model. 

\subsubsection{Injection Dynamics}
\label{sec:ICGinj}
%
To approximate the actual particle injection dynamics $f_\mathrm{inj}(t)$, we use a raised cosine function to account for the smoothness of the injection process, i.e., 
\begin{align}
    f_\mathrm{inj}(t)\approx \hat{f}_\mathrm{inj}(\mathcal{B},t) = \begin{cases} \frac{1}{t_{\mathrm{w}}} \left(1 -\cos(\omega (t - t_\mathrm{0}) \right) & t_\mathrm{0} < t < t_\mathrm{w}+t_\mathrm{0}\Hquad,\\
    0 & \text{else}\Hquad,
    \end{cases}\label{eq:raised}
\end{align}
where $\mathcal{B}=\{t_\mathrm{w}, t_{0}\}$ is the parameter set, with $t_\mathrm{w}$ representing the injection duration and defining the period of the cosine function as $\omega = 2\pi/t_{\mathrm{w}}$, and $t_0$ denoting the injection start time. 

\subsubsection{Transient and Steady State Phases}\label{subsec:ICGdisEst}
We propose two parametric models to approximate the particle distribution $I_{\mathrm{dist}}(\bm{x}_{\mathcal{R}},t)$ in \ac{ROI} $\mathcal{R}$, centered at $\bm{x}_\mathcal{R}\in\mathcal{A}$, during the \textit{transient} and \textit{steady state phases} based on the analytical model in~\eqref{eq:wrappedNormal}. 
To emphasize the dependence of $p_\mathrm{wn}(x,t)$ in~\eqref{eq:wrappedNormal} on the effective diffusion coefficient $\Deff$, effective flow velocity $\veff$, effective length $\Leff$, and also the distance $d_\mathrm{rx}$ from the injection site, we denote the distribution in \eqref{eq:wrappedNormal} in the following by $p_\mathrm{wn}(\varcal{Q},t)$, where $\mathcal{Q}$ denotes the parameter set, i.e., $\varcal{Q}=\{\Deff, \veff, \Leff, d_\mathrm{rx}\}$.


\paragraph{Single Wrapped Normal Distribution}
We first propose a basic parametric model assuming that the complete vascular system of the \ac{CAM} model can be represented by a single loop as described in Section~\ref{sec:particleDis} and shown in Fig.~\ref{fig:cam-mc-approx}. To obtain an approximation of $I_\mathrm{dist}(\bm{x}_\mathcal{R},t)$, we combine $\hat{f}_\mathrm{inj}(\mathcal{B},t)$ from \eqref{eq:raised} with the analytical channel model $p_\mathrm{wn}(\mathcal{Q},t)$ from \eqref{eq:wrappedNormal}, as follows
\begin{align}
I_{\mathrm{dist}}(\bm{x}_\varcal{R},t)\approx \hat{I}_{\mathrm{dist}}(\varcal{Q},t) =  \hat{f}_\mathrm{inj}(\mathcal{B},t) \ast p_\mathrm{wn}(\mathcal{Q},t)\Hquad,  
\label{eq:approx}
\end{align}
where $\ast$ denotes temporal convolution.

\paragraph{Multiple Wrapped Normal Distributions}\label{sec:2rings}
As discussed in Section~\ref{sec:CAM}, the \ac{CAM} model is highly vascularized. Therefore, to account for this complex branched topology, we propose an extended parametric model allowing for multiple loops, which might occur in the \ac{CAM} vascular network. Thus, we extend~\eqref{eq:approx} to the superposition of multiple wrapped normal distributions \eqref{eq:wrappedNormal} characterized by different sets of physical parameters as follows
\begin{align}
I_{\mathrm{dist}}(\bm{x}_\varcal{R},t)\approx \hat{I}_{\mathrm{dist},n}(\varcal{P}^{n},t) =  \hat{f}_\mathrm{inj}(\mathcal{B},t) \ast \sum_{j=1}^{n}a^{n,j}~p_\mathrm{wn}(\mathcal{Q}^{n,j},t)\Hquad, 
\label{eq:approx2}
\end{align}
where $a^{n,j}$ is a weight parameter, i.e., $\Sigma_{j=1}^{n}a^{n,j}=1$, and the superscript $j$ indicates the different wrapped normal distributions. The complete set of model parameters is given by $\varcal{P}^{n}=\{\varcal{P}^{n,1},\cdots,\varcal{P}^{n,n}\}$, where $\varcal{P}^{n,j} = \{a^{n,j}, \mathcal{Q}^{n,j}\} = \{a^{n,j},\Deff^{n,j}, \veff^{n,j}, \Leff^{n,j}, d_\mathrm{rx}^{n,j}\}$ denotes the parameters belonging to the $j$-th term of the summation in~\eqref{eq:approx2}. The total number of parameters is $n\times|\varcal{P}^{n,j}|$, where $|\cdot|$ is the size of a set. For $n = 1$, \eqref{eq:approx2} is equivalent to~\eqref{eq:approx}. We note that in the rest of the paper~\eqref{eq:approx} and~\eqref{eq:approx2} are referred as the basic and the extended parametric models, respectively. 

\subsubsection{Accumulation Phase}\label{sec:acc_est}

As previously described in Sections~\ref{subsec:assumpt} and \ref{sec:sec2B}, and shown in Fig.~\ref{fig:Liver}, particles are very quickly uniformly distributed in the vascular system of the \ac{CAM} model and reach a steady state before the fluorescence intensity slowly decreases in the vascular system and increases in the organs of the embryo during the \textit{accumulation phase}. The acquisition of reliable measurement data for the accumulation in the embryo's organs is very challenging, mainly due to the movement of the embryo and possible concealment of the organs by other parts of the \ac{CAM} model. 
In order to provide a comprehensive description of the molecule distribution in the \ac{CAM} model, we propose the following simple parametric model for the accumulation of molecules in the organs of the embryo during the \textit{accumulation phase}
\begin{align}
I_{\mathrm{acc}}(\bm{x}_\varcal{R},t)\approx \hat{I}_{\mathrm{acc}}(\mathcal{C},t) =  1-a\exp{(-bt)}\Hquad,  
\label{eq:accum}
\end{align}
with parameter set $\mathcal{C} = \{a, b\}$. We note that, as more measurement data for the \textit{accumulation phase} become available, we will further extend~\eqref{eq:accum} to a physically motivated model for particle accumulation in future work.
%
%

\subsection{Curve Fitting}\label{sec:CrvFit_RMSE}
After developing the parametric models in Section~\ref{sec:ParamMod}, we fit them to measured data in order to determine the parameters. For all curve fittings based on the parametric models~\eqref{eq:raised},~\eqref{eq:approx},~\eqref{eq:approx2}, and~\eqref{eq:accum}, we use Matlab's curve fitting toolbox implementing a nonlinear least squares method. The curve fitting and parameter estimation is performed based on the fluorescence intensity $I(\bm{x},t)$ measured over time with an \ac{ICG} camera focusing on the opened egg area $\varcal{A}$ with area size $A$\footnote{We assume the measured \ac{ICG} fluorescence intensity is proportional to the \ac{ICG} concentration in $\bm{x}_\varcal{R}$ \cite{Chon2023}.}. In particular, $I(\bm{x},t)$ denotes the measured fluorescence intensity of a region centered at $\bm{x}$ in the open egg area $\varcal{A}$. As the absolute numerical values of the measured  fluorescence intensity are not physically meaningful, all measured intensities are normalized to their steady state value to ensure comparability within individual measurements. We employ the \ac{RMSE} as a metric to evaluate the performance of the parametric models approximating the \ac{ICG} dynamic.

\subsubsection{Injection Dynamics}
For the injection dynamics, we first calculate the mean fluorescence intensity $\bar{I}(t)$ in the open egg area $\varcal{A}$, i.e., $\bar{I}(t) = \frac{1}{A}\int_{\varcal{A}} I(\bm{x},t)\,\mathrm{d}\bm{x}$. Then, the actual injection dynamics can be obtained from the derivative of the mean intensity $\bar{I}(t)$, i.e., $f_\mathrm{inj}(t) = \frac{\mathrm{d}}{\mathrm{d} t} \bar{I}(t)$. The derivative of the mean intensity $\bar{I}(t)$ is computed numerically as measurements are available only at discrete times $i\Delta t$, i.e., $\bar{I}(i\Delta t)$, where $\Delta t$ is the sampling interval, $i\in\{0,1,\ldots,N\}$, and $N+1$ is the total number of discrete-time samples withing the observation interval. The numerical derivative is defined as $\frac{\Delta \bar{I}(i\Delta t)}{\Delta t}= \frac{\bar{I}(i\Delta t)-\bar{I}((i-1)\Delta t)}{\Delta t}$, and $\bar{I}(-\Delta t)=0$.
Hence, by fitting the parametric model $\hat{f}_\mathrm{inj}(\mathcal{B},t)$ to the measured injection dynamics $f_\mathrm{inj}(t)$, $t_\mathrm{w}$ and $t_\mathrm{0}$ can be jointly estimated.
The nonlinear least squares curve fitting method solves the following optimization problem
\begin{equation}\label{eq:NLSfinj}
    \mathcal{B} = \arg\min_{\tilde{\mathcal{B}}\in\mathcal{S}_\mathcal{B}} \sum_{i=0}^{N}\left( f_{\mathrm{inj}}(i\Delta t)-\hat{f}_{\mathrm{inj}}(\tilde{\mathcal{B}},i\Delta t)\right)^{2}\Hquad,
\end{equation}
where the parameter set $\mathcal{B}\in\mathcal{S}_\mathcal{B}$ minimizes the objective function~\eqref{eq:NLSfinj}, and $\mathcal{S}_{\mathcal{B}}\subseteq\mathbb{R}^2$ is the parameter search space. 
\subsubsection{Transient and Steady State Phases}
For the approximation of the particle dynamics during the \textit{transient} and \textit{steady state phases}, we extract the localized fluorescence intensity $I_{\mathrm{dist}}(\bm{x}_{\varcal{R}},t)$ in a specific \ac{ROI} $\varcal{R} \subset \varcal{A}$, centered at $\bm{x}_{\varcal{R}}$. As all physical parameters of the \ac{CAM} vascular system are unknown or only given very vaguely in the literature, the parameter sets $\mathcal{Q}$ in \eqref{eq:approx} and $\mathcal{P}^n$ in \eqref{eq:approx2} are estimated during the fitting process. Moreover, as the measured $I_{\mathrm{dist}}(\bm{x}_\varcal{R},t)$ reaches steady state very quickly, the initial slope is the most important feature for the fitting process,  the actual fitting of \eqref{eq:approx} and \eqref{eq:approx2} is performed based on the derivatives of the measured fluorescence intensity, i.e., $\frac{\mathrm{d}}{\mathrm{d}t}\hat{I}_{\mathrm{dist}}(\mathcal{P}^{n},t) \approx \frac{\mathrm{d}}{\mathrm{d}t}I_{\mathrm{dist}}(\bm{x}_{\varcal{R}},t)$. Hence, the nonlinear least squares method to obtain parameter sets $\mathcal{Q}$ and $\mathcal{P}^n$ solves the following optimization problems
\begin{equation}\label{eq:MLqrtQ}
    \mathcal{Q} = \arg\min_{\tilde{\mathcal{Q}}\in\mathcal{S}_{\mathcal{Q}}} \sum_{i=0}^{N}\left(\frac{\Delta}{\Delta t} I_{\mathrm{dist}}(\bm{x}_{\varcal{R}},i\Delta t)-\frac{\Delta}{\Delta t}\hat{I}_{\mathrm{dist}}(\tilde{\mathcal{Q}},i\Delta t)\right)^{2}\Hquad,
\end{equation}

\begin{equation}\label{eq:MLqrtP}
    \mathcal{P}^n = \arg\min_{\tilde{\mathcal{P}}^{n}\in\mathcal{S}_{\mathcal{P}^n}} \sum_{i=0}^{N}\left(\frac{\Delta}{\Delta t} I_{\mathrm{dist}}(\bm{x}_{\varcal{R}},i\Delta t)-\frac{\Delta}{\Delta t}\hat{I}_{\mathrm{dist},n}(\tilde{\mathcal{P}}^n,i\Delta t)\right)^{2}\Hquad,
\end{equation}
where the optimal parameter sets $\mathcal{Q}\in\mathcal{S}_{\mathcal{Q}}$ and $\mathcal{P}^n\in\mathcal{S}_{\mathcal{P}^n}$ minimize the objective functions in~\eqref{eq:MLqrtQ} and~\eqref{eq:MLqrtP}, respectively, and $\mathcal{S}_{\mathcal{Q}}\subseteq\mathbb{R}^{4}$ and $\mathcal{S}_{\mathcal{P}^n}\subseteq\mathbb{R}^{5n}$ are the corresponding parameter search spaces.
\subsubsection{Accumulation Phase}
For the approximation of the particle dynamics during the \textit{accumulation phase}, we measure the fluorescence intensity $I_{\mathrm{acc}}(\bm{x}_{\varcal{R}},t)$. Parameter set $\mathcal{C}$ is estimated via curve fitting of $\hat{I}_{\mathrm{acc}}(\mathcal{C}, t)$ in \eqref{eq:accum} to measured fluorescence intensity during the \textit{accumulation phase}, $I_{\mathrm{acc}}(\bm{x}_\varcal{R},t)$. Numerically, the fitting process to obtain the parameter set $\mathcal{C}$ can be expressed as
\begin{equation}\label{eq:NLMSacc}
    \mathcal{C} = \arg\min_{\tilde{C}\in \mathcal{S}_{\mathcal{C}}} \sum_{i=0}^{N}\left( I_{\mathrm{acc}}(\bm{x}_{\varcal{R}},i\Delta t)-\hat{I}_{\mathrm{acc}}(\tilde{C},i\Delta t)\right)^{2}\Hquad,
\end{equation}
where the parameter set $\mathcal{C}\in\mathcal{S}_\mathcal{C}$ minimizes the objective function in \eqref{eq:NLMSacc}, and $\mathcal{S}_{\mathcal{C}}\subseteq\mathbb{R}^2$ is the parameter search space.
\subsubsection{Fitting Performance}
To evaluate the accuracy of the developed parametric models, an appropriate performance metric is crucial. As the fitting of the parametric models is applied to the entire dataset containing measurements of the \textit{transient} and \textit{steady state phases}, the data might contain additional noise and potentially unknown influencing factors. Thus, we choose the \ac{RMSE} as performance metric due to its robustness to noisy data. The \ac{RMSE} between a measurement $I_{\mathrm{X}}(\bm{x}_{\varcal{R}},t)$ and its approximation $\hat{I}_{\mathrm{X}}(\mathcal{V},t)$ is defined as follows
\begin{align}
\text{RMSE}\left(\hat{I}_{\mathrm{X}}\right) = \sqrt{\frac{1}{N}\sum^{N}_{i=0}\left(I_{\mathrm{X}}(\bm{x}_{\varcal{R}},i\Delta t) - \hat{I}_{\mathrm{X}}(\mathcal{V},i\Delta t) \right)^2}\Hquad,
\label{eq:RMSE}
\end{align}
where $\mathrm{X} = \{\mathrm{dist},\,, (\mathrm{dist},n), \, \mathrm{acc}\}$ and $\mathcal{V} = \{\mathcal{Q},\,\mathcal{P}^n,\, \mathcal{C} \}$ are placeholders to distinguish between the parametric models in \eqref{eq:approx}, \eqref{eq:approx2}, and \eqref{eq:accum}, respectively. 
In the following section, we will use the \ac{RMSE} in \eqref{eq:RMSE} to evaluate the accuracy of the parametric models for the \textit{transient} and \textit{steady state phases} in \eqref{eq:approx} and \eqref{eq:approx2} and the \textit{accumulation phase} in \eqref{eq:accum}. 

\section{Experimental ICG Distribution Study}
\label{sec:exp}
In this section, we present experimental and analytical results for the distribution of \ac{ICG} molecules in the \ac{CAM} model. Similar to the previous sections, we discuss the \textit{transient} and \textit{steady state phases} and the \textit{accumulation phase} separately.
First, we consider the \textit{transient} and \textit{steady state phases}. We briefly describe the employed measurement methods, and the prepared dataset containing the measurement data, estimated parameters, and approximated \ac{ICG} distribution~\cite{ICGdistDataset2025}. Then, we evaluate the approximation of the \ac{ICG} injection dynamics by \eqref{eq:raised} exemplarily for three measurements contained in the dataset. Subsequently, the approximation of the \ac{ICG} distribution in the \textit{transient} and \textit{steady state phases} based on~\eqref{eq:approx} and~\eqref{eq:approx2} is evaluated. Furthermore, we present the estimated parameters obtained for both parametric models and discuss their plausibility and consistency. 
Second, we study the \textit{accumulation phase}. We describe the corresponding measurement process and related challenges. Then, we provide first measurement results for two eggs and compare with the results obtained for the parametric model~\eqref{eq:accum}. 

\subsection{Transient and Steady State Phases}

\subsubsection{Measurement Process and Dataset Description}
For all measurements, we injected the fluorescent molecule \ac{ICG} into the \ac{CAM} vascular system via a syringe and the fluorescence intensity $I(\bm{x}_\mathcal{R},t)$ was tracked by an \ac{ICG} camera focusing on the open egg area $\mathcal{A}$. From the recordings, the \ac{ICG} intensities $I(\bm{x}_\mathcal{R},t)$ in different \acp{ROI} centered at $\bm{x}_\mathcal{R}\in\mathcal{A}$ were extracted and post-processed. A detailed description of the experimental preparation of the eggs can be found in the Appendix~\ref{sec:methods}. 
For the generated dataset, we performed experiments on $35$ different eggs at different \acp{DED}, where the measurements of ten eggs had to be excluded. We measured $125$ \ac{ROI}s on the $25$ remaining eggs and obtained $69$ reliable measurements. The two main reason that caused the loss of eggs were bleeding after the \ac{ICG} injection and the embryo's movement interfering with the measurements. In particular, after the removal of the injection needle, blood can spread on the surface of the \ac{CAM} and distort the measurements. The embryo's movement results in unpredictable changes in the position of the vessels that were marked as \ac{ROI}, which made a reliable tracking of the \ac{ICG} intensity impossible.
Based on the data successfully collected from 25 eggs we created a dataset for the particle distribution in the \ac{CAM} model during the \textit{transient} and \textit{steady state phases}, which can be downloaded from~\cite{ICGdistDataset2025}. 
The dataset is a~\textit{.mat} file and named \textit{CAM\_Data.mat}. This file contains all the data in a single variable of \textit{table} type named \textit{MetaData}. The \textit{MetaData} variable contains the egg numbers, ROI numbers, parameter sets $\mathcal{P}^1$ and $\mathcal{P}^2$, \ac{DED}, injection duration, $I_{\mathrm{dist}}(\bm{x}_\mathcal{R},t)$, time vector, approximate parametric models $\hat{I}_{\mathrm{dist},1}(\mathcal{P}^1,t)$, $\hat{I}_{\mathrm{dist},2}(\mathcal{P}^2,t)$, images of the CAM, $\text{RMSE}(\hat{I}_{\mathrm{dist},1})$, $\text{RMSE}(\hat{I}_{\mathrm{dist},2})$, and injection distance $d_{\mathrm{inj}}$, which is the distance between the injection point and the ROI of interest (e.g., the white double arrow in the CAM model image of Egg 20 in Fig.~\ref{fig:2Ring}). 

\subsubsection{Molecule Injection Dynamics}
\begin{figure}[t!]
    \centering
    \includegraphics[width=0.8\linewidth]{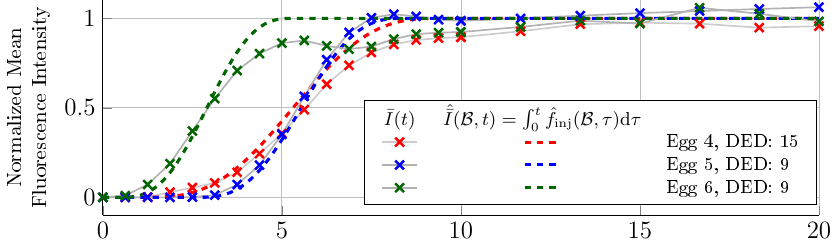}\\
    \includegraphics[width=0.8\linewidth]{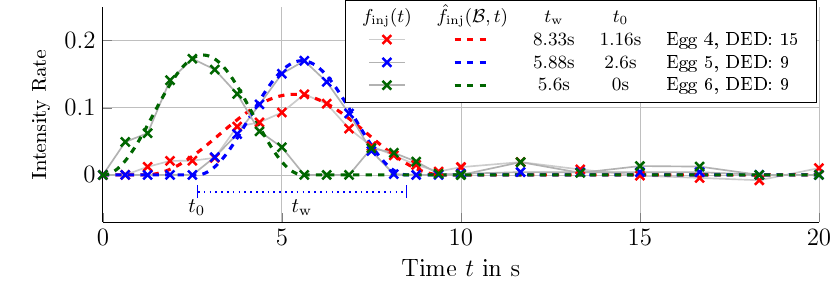}
    \caption{\small Top: Measured $\bar{I}(t)$ and fitted $\hat{\bar{I}}(t)$ mean fluorescence intensity for three eggs. Bottom: Measured and fitted injection dynamics $\hat f_\mathrm{inj}(t)$ and estimated injection delays and duration.}
    \label{fig:injection_est}
\end{figure}
The curves with markers in Fig.~\ref{fig:injection_est} show the measured mean fluorescence intensity $\bar{I}(t)=\frac{1}{A}\int_\mathcal{A}I(\bm{x},t)\mathrm{d}\bm{x}$ (top plot) and the injection dynamics $f_\mathrm{inj}(t) = \frac{\mathrm{d}}{\mathrm{d}t}\bar{I}(t)$ (bottom plot) for three different eggs from the dataset. 
The dashed curves in the bottom plot of Fig.~\ref{fig:injection_est} show the approximation $\hat{f}_\mathrm{inj}(\mathcal{B},t)$ of $f_\mathrm{inj}(t)$ according to \eqref{eq:raised} obtained by curve fitting alongside the estimated injection time $t_0$ and duration $t_\mathrm{w}$. The resulting approximated mean intensity $\hat{\bar{I}}(\mathcal{B},t) = \int_0^t \hat{f}_\mathrm{inj}(\mathcal{B},\tau)\mathrm{d}\tau$ is shown by the dashed curves in the top plot of Fig.~\ref{fig:injection_est}. From both plots, it can be observed that the approximation $\hat{f}_\mathrm{inj}(\mathcal{B},t)$ obtained with the simple model \eqref{eq:raised} is in good agreement with the measured injection dynamics $f_\mathrm{inj}(t)$ for all considered eggs. 

\begin{figure}[h!]
    \centering
    \includegraphics[width=0.9\linewidth]{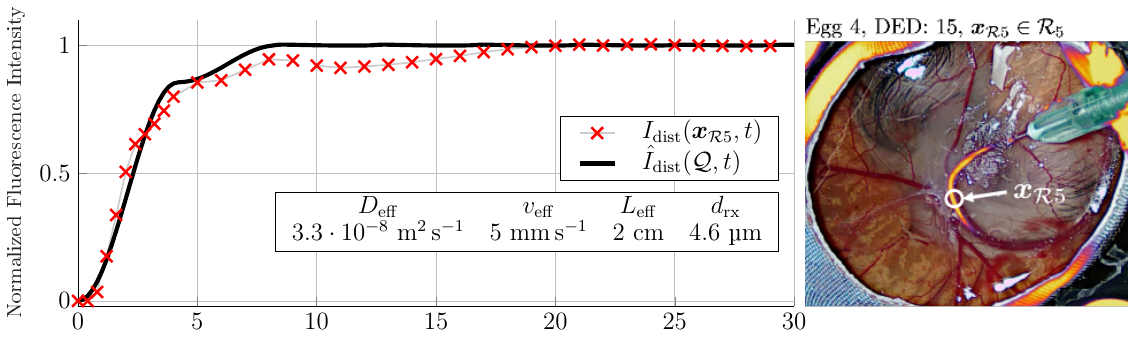}\\[-0.5em]
    \includegraphics[width=0.9\linewidth]{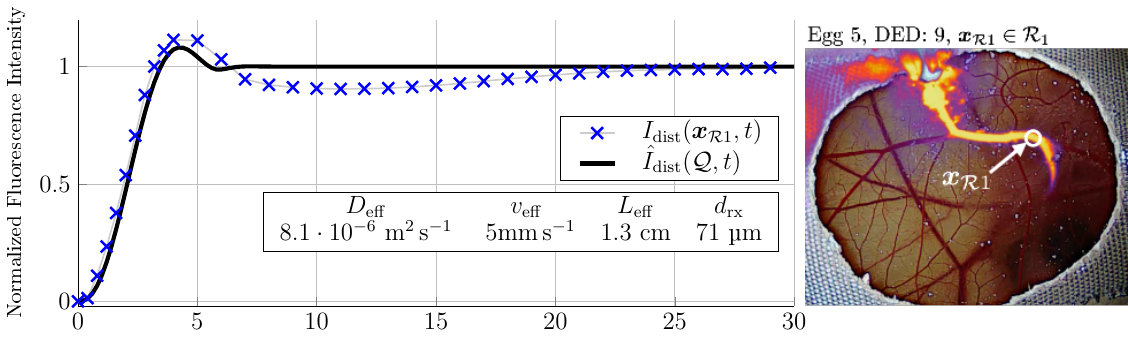}\\[-0.5em]
    \includegraphics[width=0.9\linewidth]{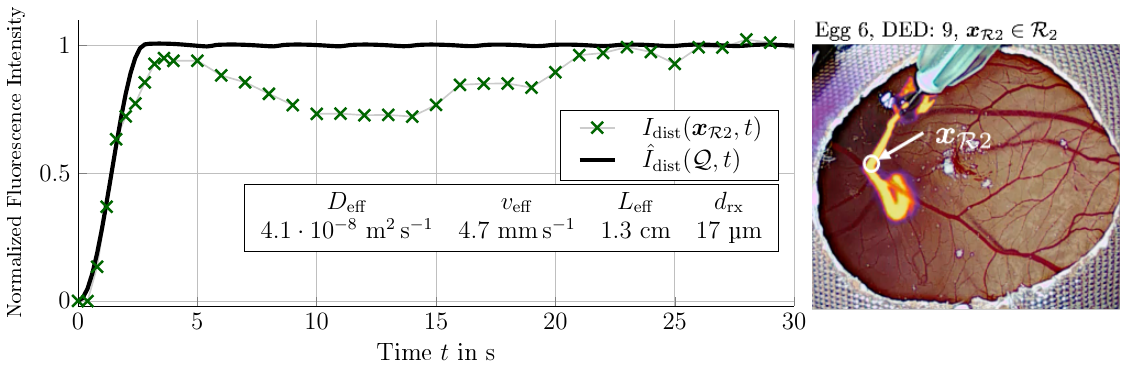}
    \caption{\small Left: Three chosen measurements of the \ac{ICG} distribution $I_{\mathrm{dist}}(\bm{x}_\varcal{R},t)$  from the dataset (curves with markers) and approximations $\hat{I}_\mathrm{dist}(\varcal{Q},t)$ (solid curves) obtained with the basic parametric model in \eqref{eq:approx}. Right: Photo of the considered eggs shortly after the injection. The specific \ac{ROI} positions, $\bm{x}_{\mathcal{R}5}$, $\bm{x}_{\mathcal{R}1}$, and $\bm{x}_{\mathcal{R}2}$, are highlighted by a circle and an arrow. The numbers indicate the number of the \ac{ROI} in a specific egg, e.g., $\mathcal{R}_5$ in Egg $4$ is the fifth considered \ac{ROI} where measurement data were obtained.}
    \label{fig:results}
\end{figure}

\subsubsection{Molecule Distribution in Transient and Steady State Phases}
\paragraph{Single Wrapped Normal Distribution}
In the following, we evaluate the capability of the basic parametric model~\eqref{eq:approx} to approximate the \textit{transient} and \textit{steady state phases} of molecule distribution in the \ac{CAM} model (see Fig.~\ref{fig:cam-mc-approx} and Section~\ref{sec:particleDis}). 
Figure~\ref{fig:results} shows the measured fluorescence intensity $I_\mathrm{dist}(\bm{x}_\varcal{R},t)$ at $\bm{x}_\varcal{R}$ over time, for three eggs from the dataset (curves with colored markers). The black curves in Fig.~\ref{fig:results} show the approximated fluorescence intensity $\hat{I}_\mathrm{dist}(\mathcal{Q},t)$ obtained by curve fitting of \eqref{eq:approx} and the estimated parameter values for each measurement. The photos on the right hand side of Fig.~\ref{fig:results} show the actual injection locations and the locations of \acp{ROI} $\varcal{R}_5$, $\varcal{R}_1$, and $\varcal{R}_2$ for Egg 4, Egg 5, and Egg 6, respectively, shortly after the injection started. In general, all measurements are in line with the assumption that the \textit{steady state phase} inside the vascular system is reached very quickly (see Section~\ref{subsec:assumpt}). 
We observe from Fig.~\ref{fig:results} that the overall shape of the measured $I_\mathrm{dist}(\bm{x}_\varcal{R},t)$ is well reproduced by the proposed parametric model $\hat{I}_\mathrm{dist}(\mathcal{Q},t)$. For example, for Egg 4 (top plot) the characteristic increases in the slope around $5~\si{\second}$, for Egg 5 (center plot) the peak around $4~\si{\second}$, and for Egg 6 (bottom plot) the initial slope are reproduced very well by the proposed model. However, for all eggs considered in Fig.~\ref{fig:results}, the basic parametric model~\eqref{eq:approx} fails to capture the intensity drop (around $10~\si{\second}$), most prominently for Egg 6 in the bottom plot. Instead, approximation $\hat{I}_\mathrm{dist}(\mathcal{Q},t)$ stays at the steady state. Despite this mismatch, the proposed basic parametric model in~\eqref{eq:approx} is capable of approximating the complex behavior of the \ac{CAM}s vascular system for a fixed injection-measurement (\ac{TX}-\ac{RX}) arrangement. 

Next, we discuss the estimates obtained for the parameters $\mathcal{Q}$ of the basic model, which are also given in Fig.~\ref{fig:results} for each egg considered. First, we can observe that the estimated value of $\veff$ is similar for all three eggs, and is in the range of values known for the \ac{CAM} model from the literature (see Section~\ref{sec:topol}). The estimated $\Deff$ values are in the range of $10^{-8}$--$10^{-6}~\si{\square\meter\per\second}$, which exceeds the typical range of molecule diffusion coefficients by several orders of magnitude. However, as the \ac{CAM} system is highly vascularized and contains several sources of turbulence (cf. Section~\ref{sec:topol}), these large values are reasonable and reflect the expected dispersion, which is captured by the effective diffusion coefficient $\Deff$. 
Although the estimates obtained for the parameters of $\hat{I}_\mathrm{dist}(\varcal{Q},t) $ are not estimates of actual physical parameters in the \ac{CAM} model because of the high level of abstraction of the analytical and parametric models, the relationships between them seem meaningful. 

\paragraph{Multiple Wrapped Normal Distributions}\label{sec:2ringsExp}
\begin{figure*}[t!]
    \centering
    \begin{minipage}{0.49\linewidth}
        \includegraphics[width=\linewidth]{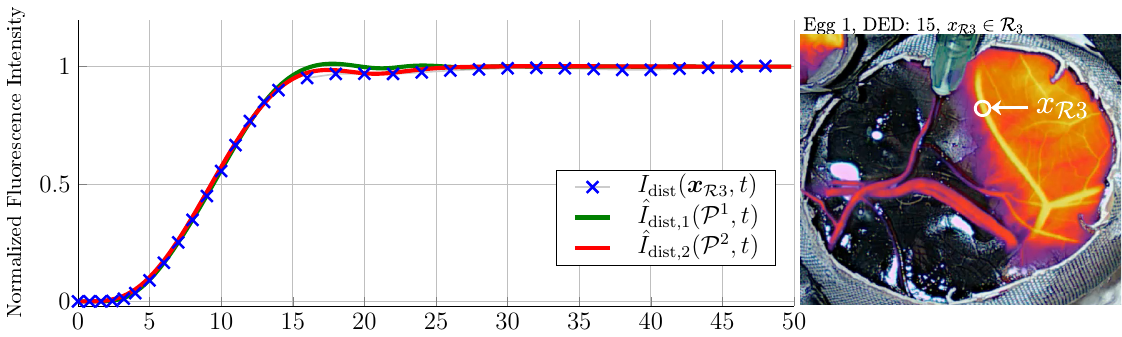}
    \end{minipage}\hfill
    \begin{minipage}{0.49\linewidth}
        \includegraphics[width=\linewidth]{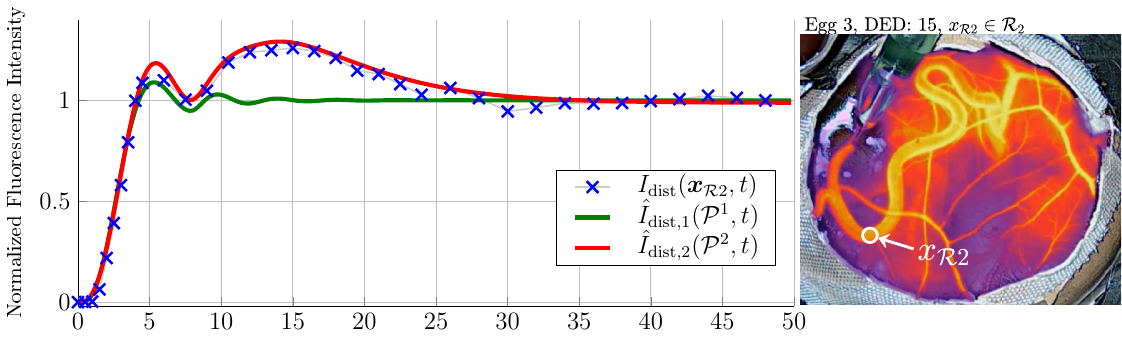}
    \end{minipage}\\
    \centering
    \begin{minipage}{0.49\linewidth}
        \includegraphics[width=\linewidth]{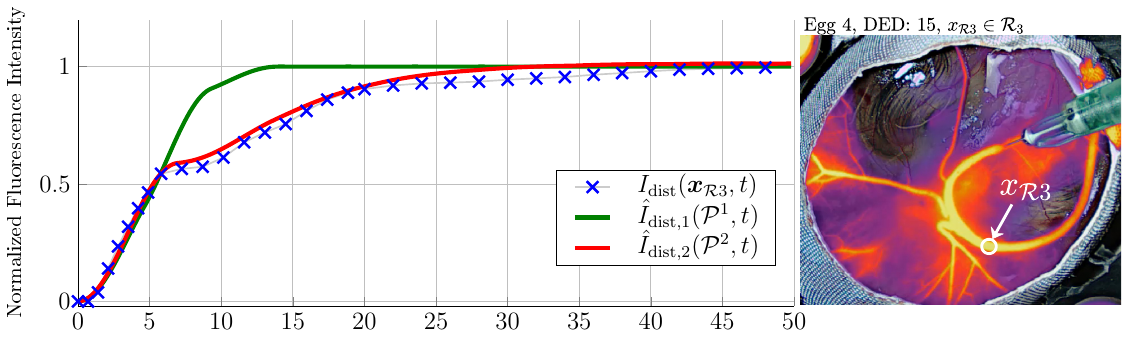}
    \end{minipage}\hfill
    \begin{minipage}{0.49\linewidth}        \includegraphics[width=\linewidth]{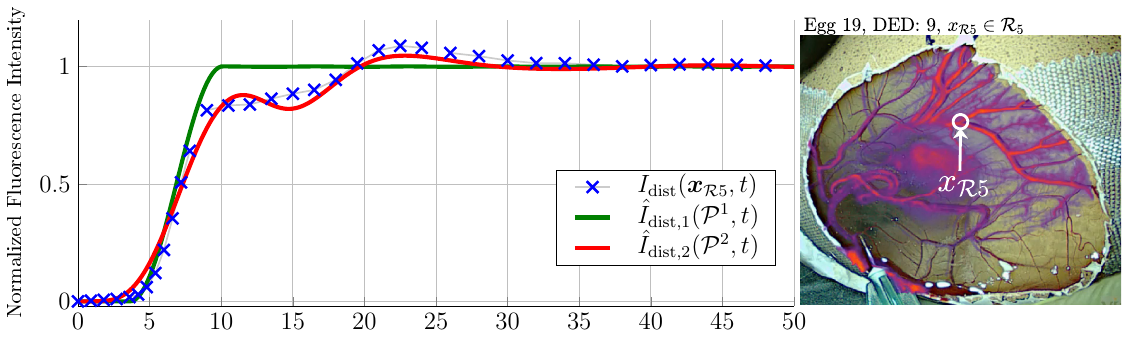}
    \end{minipage}\\
    \centering
    \begin{minipage}{0.49\linewidth}
        \includegraphics[width=\linewidth]{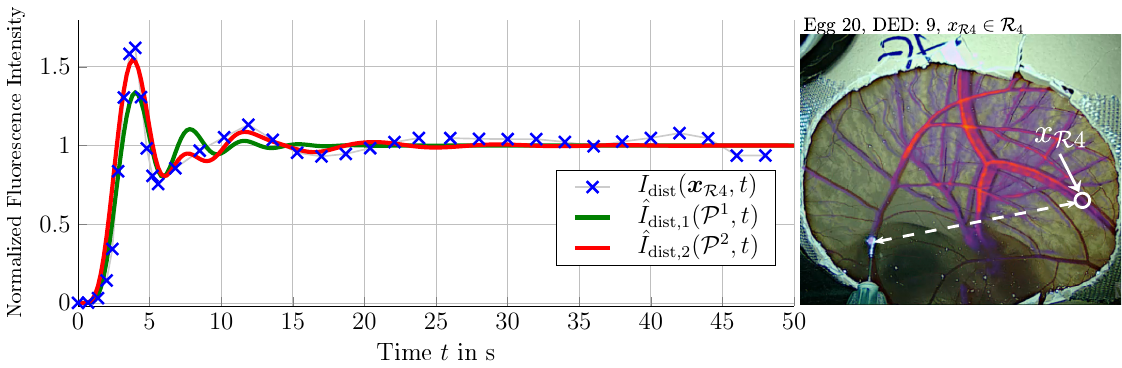}
    \end{minipage}\hfill\hspace{0.2cm}
    \begin{minipage}{0.49\linewidth}
        \includegraphics[width=\linewidth]{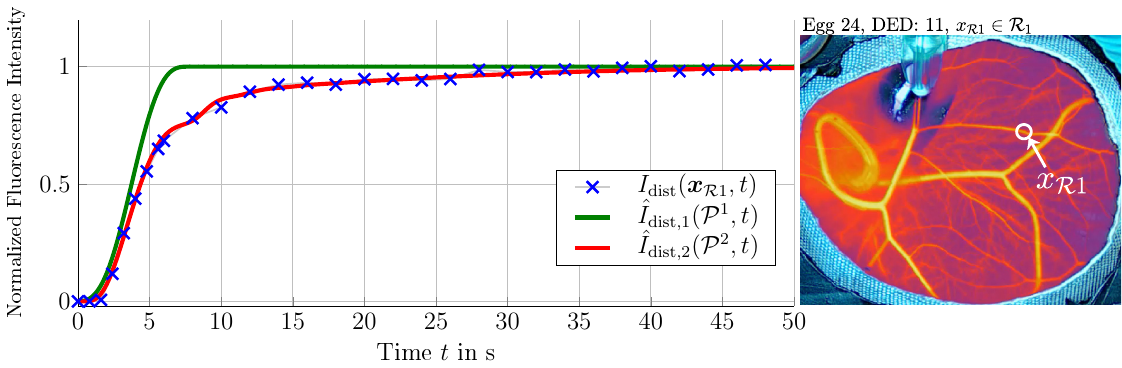}
    \end{minipage}
    \caption{\small Experimental and fitting results for parametric models $\hat{I}_{\mathrm{dist},1}(\mathcal{P}^1,t)$ and $\hat{I}_{\mathrm{dist},2}(\mathcal{P}^2,t)$ for approximating the propagation of \ac{ICG} molecules in the \ac{CAM} vascular system, measured in specific \acp{ROI} for six different eggs. An example of the injection distance, $d_{\mathrm{inj}}$, which indicates the distance between the injection point and the desired \ac{ROI}, is shown with a white dashed double arrow in the image of Egg~20.}
    \label{fig:2Ring}
\end{figure*}
\begin{table*}[t!]
\begin{center}
\caption{Estimated parameters for Fig.~\ref{fig:2Ring}.}
    \label{tab:params}
\begin{tabular}{|c|c|c|c|c|c|c|c|}
\hline
                                         & DED & $n$ & $D_{\mathrm{eff}}~\left[\si{\square\meter\per\second}\right]$ & $v_{\mathrm{eff}}~\left[\si{\milli\meter\per\second}\right]$ & $L_{\mathrm{eff}}~\left[\si{\centi\meter}\right]$ & $d_{\mathrm{rx}}~\left[\si{\centi\meter}\right]$ & $a$           \\
\hline
\multirow{2}{*}{Egg~1, $x_{\mathcal{R}3}$} & \multirow{2}{*}{15} & $1$ & $13.8\cdot 10^{-6}$                              & $9.4$                                           & $5.6$                                & $2.1$                               & $1$           \\
                                          & & $2$ & $(7.6\cdot 10^{-6},9.7\cdot 10^{-7})$            & $(2.9, 0.4)$                                    & $(2.6, 1.3)$                         & $(0.64,1)$                          & $(0.88,0.12)$ \\
\hline
\multirow{2}{*}{Egg~3, $x_{\mathcal{R}2}$} & \multirow{2}{*}{15} & $1$ & $12\cdot 10^{-6}$                               & $9.9$                                           & $4.4$                                & $4.1$                               & $1$           \\
                                          & & $2$ & $(5.2\cdot 10^{-5},4.5\cdot 10^{-8})$            & $(11, 0.2)$                                    & $(5.8, 6.1)$                         & $(0.4,0.2)$                         & $(0.98,0.02)$ \\
\hline
\multirow{2}{*}{Egg~4, $x_{\mathcal{R}3}$} & \multirow{2}{*}{15} & $1$ & $1.8\cdot 10^{-8}$                               & $8.9$                                           & $4.1$                                & $0.0006$                            & $1$           \\
                                           & & $2$ & $(3.4\cdot 10^{-8},3.1\cdot 10^{-6})$            & $(10, 0.04)$                                   & $(7, 5.2)$                           & $(0.001,1.4)$                       & $(0.65,0.35)$ \\
\hline
\multirow{2}{*}{Egg~19, $x_{\mathcal{R}5}$}& \multirow{2}{*}{9} & $1$ & $1\cdot 10^{-11}$                                & $4.3$                                           & $2.8$                                & $1.5$                               & $1$           \\
                                          & & $2$ & $(3.4\cdot 10^{-6},2.2\cdot 10^{-7})$            & $(2.2, 0.5)$                                    & $(3.6, 0.8)$                           & $(0.8,0.7)$                         & $(0.39,0.61)$ \\
\hline
\multirow{2}{*}{Egg~20, $x_{\mathcal{R}4}$}& \multirow{2}{*}{9} & $1$ & $5.3\cdot 10^{-6}$                               & $6.7$                                           & $2.5$                                & $0.8$                               & $1$           \\
                                          & & $2$ & $(3.9\cdot 10^{-6},7.3\cdot 10^{-6})$            & $(5.5, 5.3)$                                    & $(2, 4.6)$                           & $(0.6,0.5)$                         & $(0.83,0.17)$ \\
\hline
\multirow{2}{*}{Egg~24, $x_{\mathcal{R}1}$}& \multirow{2}{*}{11} & $1$ & $6.8\cdot 10^{-10}$                              & $9.6$                                           & $0.4$                                & $0.0007$                            & $1$           \\
                                          & & $2$ & $(1.3\cdot 10^{-5},2.4\cdot 10^{-7})$            & $(10, 0.03)$                                    & $(3.4, 0.9)$                         & $(1,0.7)$                           & $(0.6,0.4)$   \\
\hline
\end{tabular}
\end{center}
\end{table*}
Next, we investigate the extended parametric model~\eqref{eq:approx2} for $n=2$ and compare it with the basic model in~\eqref{eq:approx}.
Figure~\ref{fig:2Ring} shows the fluorescence intensity $I_\mathrm{dist}(\bm{x}_\varcal{R},t)$ measured at $\bm{x}_\varcal{R}$ over time, for six different eggs from the dataset (curves with blue markers). The green and red curves in Fig.~\ref{fig:2Ring} show the approximations obtained by fitting the parametric models $\hat{I}_{\mathrm{dist},1}(\varcal{P}^{1},t)$ and $\hat{I}_{\mathrm{dist},2}(\varcal{P}^2,t)$ according to~\eqref{eq:approx} and~\eqref{eq:approx2}, respectively. 

The photos on the right hand side of each plot show the injection and considered \ac{ROI} locations (shown as white circle highlighted with an arrow) $\varcal{R}_3$, $\varcal{R}_2$, $\varcal{R}_3$, $\varcal{R}_5$, $\varcal{R}_4$, and $\varcal{R}_1$ for Egg 1, Egg 3, Egg 4, Egg 19, Egg 20, and Egg 24, respectively, shortly after the injection started. Similar to Fig.~\ref{fig:results}, all measurements shown in Fig.~\ref{fig:2Ring} are in line with the assumption that the \textit{steady state phase} inside the vascular system is reached very quickly (see Section~\ref{subsec:assumpt}). Comparing the approximations obtained by the parametric models, we can observe that the extended model $\hat{I}_{\mathrm{dist},2}(\varcal{P}^{2},t)$ is in very good agreement with the measurements and provides a much higher accuracy compared to the basic model $\hat{I}_{\mathrm{dist},1}(\varcal{P}^{1},t)$. For all six eggs, both models are able to capture the dynamics of the rapid increase of the \ac{ICG} intensity in the first few seconds. However, the basic model $\hat{I}_{\mathrm{dist},1}(\varcal{P}^{1},t)$ fails to mimic complex dynamics, as they occur, e.g., for Egg 3. However, for Egg 1 (top left plot), there is no clear peak and drop in the measured signal and both models are able to accurately approximate $I_{\mathrm{dist}}(\bm{x}_{\mathcal{R}3},t)$. For the remaining eggs, $\hat{I}_{\mathrm{dist},1}(\varcal{P}^{1},t)$ fails to capture the exact dynamics of $I_\mathrm{dist}(\bm{x}_\varcal{R},t)$ for $t>5~\si{\second}$ until the \textit{steady state phase} is reached. For Egg 20 (bottom left plot), $\hat{I}_{\mathrm{dist},1}(\varcal{P}^{1},t)$ approximates some of the characteristic peaks of $I_{\mathrm{dist},1}(\bm{x}_{\varcal{R}4},t)$ around $4.5~\si{\second}$, $8~\si{\second}$, and $11~\si{\second}$, but is not able to capture them completely. In summary, it can be observed from Fig.~\ref{fig:2Ring} that the extended parametric model in \eqref{eq:approx2} provides a better approximation of the distribution dynamics of molecules in the \ac{CAM} model during the \textit{transient} and \textit{steady state phases} than the basic model in \eqref{eq:approx}. We further investigate the accuracy of both proposed parametric models for the entire dataset in Section~\ref{subce:dataset-eval}.  

For the measurements considered in Fig.~\ref{fig:2Ring}, the estimates obtained for the parameters of $\hat{I}_{\mathrm{dist},1}(\varcal{P}^1,t)$ and $\hat{I}_{\mathrm{dist},2}(\varcal{P}^2,t)$ are shown in Table~\ref{tab:params}. For $n=2$, we show the estimated parameters of the same type as pairs (e.g., $(\Deff^{2,1}, \Deff^{2,2})$). Inspecting both Fig.~\ref{fig:2Ring} and Table~\ref{tab:params}, there are two main observations. First, most of the estimated parameters are consistent in terms of the range of the estimated parameters. 
Second, there is a clear relationship between the presence of outliers in the estimated parameters for the basic model, $\hat{I}_{\mathrm{dist},1}(\varcal{P}^{1},t)$, and its performance to capture the dynamics of the measurements in certain eggs, compared to the extended model, $\hat{I}_{\mathrm{dist},2}(\varcal{P}^{2},t)$.
In Fig.~\ref{fig:2Ring}, we observe that for Egg 4, Egg 19, and Egg 24, $\hat{I}_{\mathrm{dist},1}(\varcal{P}^{1},t)$ fails to capture the dynamics of the measurements. Accordingly, in Table~\ref{tab:params}, we observe that there are outliers in the corresponding estimated parameter sets for $\hat{I}_{\mathrm{dist},1}(\varcal{P}^{1},t)$. For example, for Egg 19, the estimated effective diffusion coefficient, $\Deff^{1,1}$, is $10^{-11}~\si{\square\meter\per\second}$, which is much smaller compared to the other estimated $\Deff$ values. Also, for Egg 4 and Egg 24, the same comparably small values of $\Deff$ were obtained. 
Moreover, for Egg 4 and Egg 24, distance $d^{1,1}_\mathrm{rx}$ is estimated as $6~\si{\micro\meter}$ and $7~\si{\micro\meter}$, respectively, which is much smaller compared to the other estimated $d_\mathrm{rx}$ values. 

\begin{figure}[t!]
    \centering
    \includegraphics[width=0.7\linewidth]{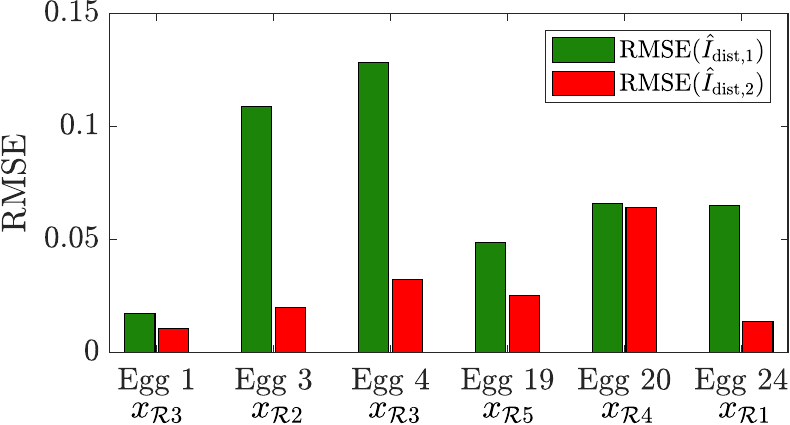}
    \caption{\small \ac{RMSE} values calculated according to~\eqref{eq:RMSE} for the approximations of $I_\mathrm{dist}(\bm{x}_\varcal{R},t)$ by the basic parametric model $\hat{I}_{\mathrm{dist},1}(\varcal{P}^{1},t)$ (green bars) in \eqref{eq:approx} and the extended parametric model $\hat{I}_{\mathrm{dist},2}(\varcal{P}^{2},t)$ (red bars) in \eqref{eq:approx2}, for the eggs considered in Fig.~\ref{fig:2Ring}.}
    \label{fig:Bar}
\end{figure}

Figure~\ref{fig:Bar} shows the \ac{RMSE} between the measurement data of the \ac{ICG} distribution and  their approximations based on the proposed parametric models for the eggs considered in Fig.~\ref{fig:2Ring}. The horizontal axis in Fig.~\ref{fig:Bar} shows the \ac{ROI} and egg number. The green and red bars show the \ac{RMSE} values according to \eqref{eq:RMSE} for the approximations of the measurements $I_{\mathrm{dist}}(\bm{x}_\varcal{R},t)$ by the basic model, $\text{RMSE}\left(\hat{I}_{\mathrm{dist},1}\right)$, and the extended parametric model, $\text{RMSE}\left(\hat{I}_{\mathrm{dist},2}\right)$, respectively. From Fig.~\ref{fig:Bar}, we observe that the accuracy of the approximation observed in Fig.~\ref{fig:2Ring} is consistent with the \ac{RMSE} values. For example, for Egg 1, Fig.~\ref{fig:2Ring} suggests that both parametric models provide a good approximation of the measurement data, resulting in very low \ac{RMSE} values in Fig.~\ref{fig:Bar}. In contrast, for Egg 3 and Egg 4, the basic parametric model $\hat{I}_{\mathrm{dist},1}(\varcal{P}^{1},t)$ yields significantly higher \ac{RMSE} values compared to the extended model $\hat{I}_{\mathrm{dist},2}(\varcal{P}^{2},t)$. This difference arises from the mismatch between the approximation by the basic model and the measurement data, which is evident in Fig.~\ref{fig:2Ring}. For Egg 20, the RMSE values of both approximations are comparable, which is also evident from Fig.~\ref{fig:2Ring}, as both models capture the dynamics with good accuracy. In conclusion, the analysis of the \ac{RMSE} values in Fig.~\ref{fig:Bar} confirms that the extended model consistently outperforms the basic model for all eggs considered in Fig.~\ref{fig:2Ring}.
\subsubsection{Estimated Parameters and RMSE of the Dataset}\label{subce:dataset-eval}
\begin{figure*}[t!]
    \centering
    \includegraphics[width=1\linewidth]{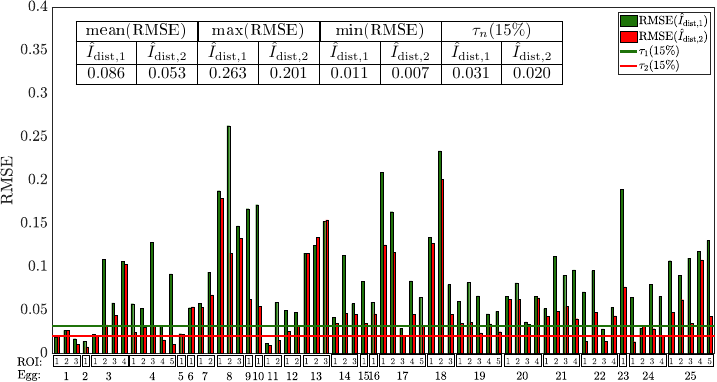}
    \caption{\small RMSE values for the complete dataset calculated according to \eqref{eq:RMSE} for the approximation of $I_\mathrm{dist}(\bm{x}_\mathcal{R},t)$ by the basic parametric model $\hat{I}_{\mathrm{dist},1}(\varcal{P}^{1},t)$ (green bars) in \eqref{eq:approx}, and the extended model $\hat{I}_{\mathrm{dist},2}(\varcal{P}^{2},t)$ (red bars) in \eqref{eq:approx2}, respectively. The horizontal green and red lines are the thresholds to indicate the best $15\%$ of the approximations. The table in the top left shows the mean, minimum, maximum, and $\tau_n (15\%)$ of the RMSEs corresponding to the curve fitting of $\hat{I}_{\mathrm{dist},1}(\varcal{P}^{1},t)$ and $\hat{I}_{\mathrm{dist},2}(\varcal{P}^{2},t)$ to $I_{\mathrm{dist}}(\bm{x}_\mathcal{R},t)$.}
    \label{fig:ApxBar}
\end{figure*}

In this section, we analyze the complete dataset of measurements of the particle distribution in the \ac{CAM} model (see \cite{ICGdistDataset2025}) along with the parameters estimated via curve fitting of the parametric models described in Section~\ref{subsec:ICGdisEst}. 
In particular, we investigate the accuracy of the approximations obtained for the basic~\eqref{eq:approx} and the extended~\eqref{eq:approx2} parametric models. Figure~\ref{fig:ApxBar} shows the \ac{RMSE} values \eqref{eq:RMSE} for the approximation of the measurements $I_\mathrm{dist}(\bm{x}_\mathcal{R},t)$ by the basic and extended models for all 69 \acp{ROI} in the dataset. On the horizontal axis, the egg numbers and \acp{ROI} are shown, and the vertical axis provides the corresponding \ac{RMSE} values. The green and red bars represent the \ac{RMSE} values obtained for the basic parametric model, $\hat{I}_{\mathrm{dist},1}(\varcal{P}^{1},t)$ in \eqref{eq:approx}, and the extended parametric model, $\hat{I}_{\mathrm{dist},2}(\varcal{P}^{2},t)$ in \eqref{eq:approx2}, respectively. The green and red horizontal lines are the thresholds to characterizing the best 15\% of the $\hat{I}_{\mathrm{dist},1}(\varcal{P}^{1},t)$ and $\hat{I}_{\mathrm{dist},2}(\varcal{P}^{2},t)$ approximations based on the \ac{RMSE} values. These thresholds are denoted by $\tau_{n}(15\%)$, $n\in\{1,2\}$. We note that, due to the high variability in the estimated parameters, we decided to not rely on the mean or median as representative values. Instead, we tightened the selection criteria and defined ``good approximations'' as those having the lowest $15\%$ of RMSE values.
The table in the top left corner of Fig.~\ref{fig:ApxBar} summarizes the performance of $\hat{I}_{\mathrm{dist},1}(\varcal{P}^{1},t)$ and $\hat{I}_{\mathrm{dist},2}(\varcal{P}^{2},t)$ across all ROIs in the dataset, by the mean, minimum, and maximum \ac{RMSE} values. Additionally, the table presents the threshold values $\tau_n(15\%)$ for both $\hat{I}_{\mathrm{dist},1}(\varcal{P}^{1},t)$ and $\hat{I}_{\mathrm{dist},2}(\varcal{P}^{2},t)$.

From Fig.~\ref{fig:ApxBar}, we can observe that the $\text{RMSE}(\hat{I}_{\mathrm{dist},2})$ values obtained for the extended model (red bars) are always smaller than the $\text{RMSE}(\hat{I}_{\mathrm{dist},1})$ values for the basic model (green bars). This confirms the observation previously made in Fig.~\ref{fig:Bar} that the approximation based on the extended model \eqref{eq:approx2} outperforms that based on the basic model \eqref{eq:approx} for the complete dataset.
%
However, it can be also observed that in some cases the basic model provides a similar accuracy as the extended model (e.g., Egg 1, Egg 2, Egg 11). The accuracy heavily depends on the egg, considered \ac{ROI}, and foremost on the dynamics of the particle propagation. For example, the dynamics of the measurement in Egg 1 in \ac{ROI}~3, top left plot of Fig.~\ref{fig:2Ring}, are very smooth, and both models provide a very accurate fit, yielding similarly low \ac{RMSE} values. 

\begin{figure*}[t]
    \centering
    \begin{minipage}{0.49\linewidth}
        \includegraphics[width=\linewidth]{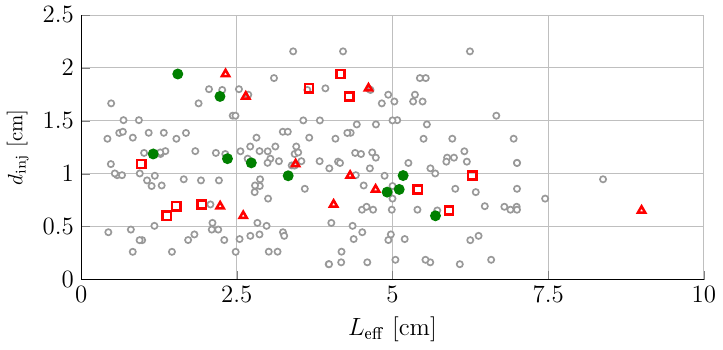}\vspace*{-0.5cm}\caption*{(a)}
    \end{minipage}\hfill
    \begin{minipage}{0.49\linewidth}
        \includegraphics[width=\linewidth]{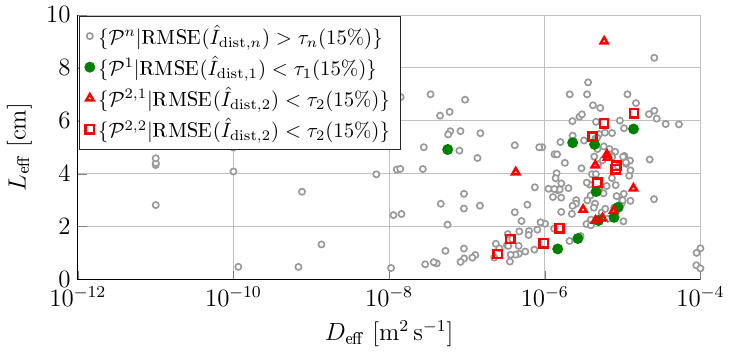}\vspace*{-0.5cm}\caption*{(b)}
    \end{minipage}\\
    \centering
    \begin{minipage}{0.49\linewidth}
        \includegraphics[width=\linewidth]{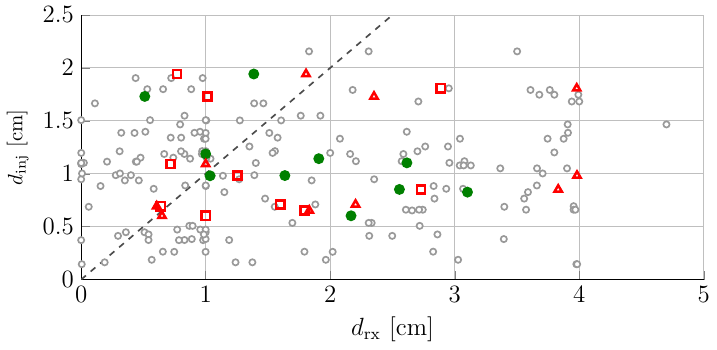}\vspace*{-0.5cm}\caption*{(c)}
    \end{minipage}\hfill
    \begin{minipage}{0.49\linewidth}
        \includegraphics[width=\linewidth]{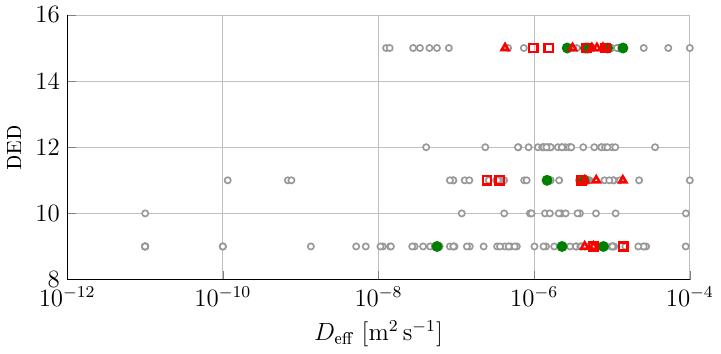}\vspace*{-0.5cm}\caption*{(d)}
    \end{minipage}\\
    \centering
    \begin{minipage}{0.49\linewidth}
        \includegraphics[width=\linewidth]{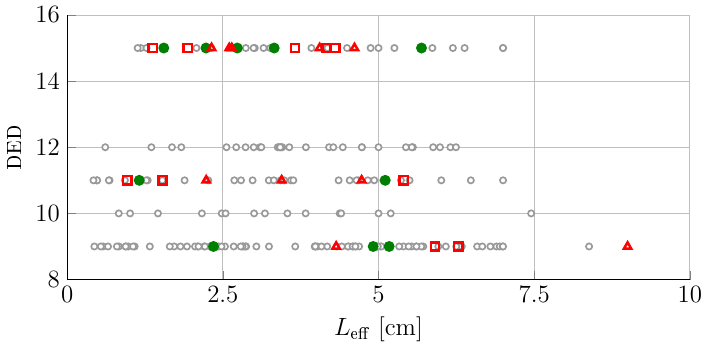}\vspace*{-0.5cm}\caption*{(e)}
    \end{minipage}\hfill\hspace{0.2cm}
    \begin{minipage}{0.49\linewidth}
        \includegraphics[width=\linewidth]{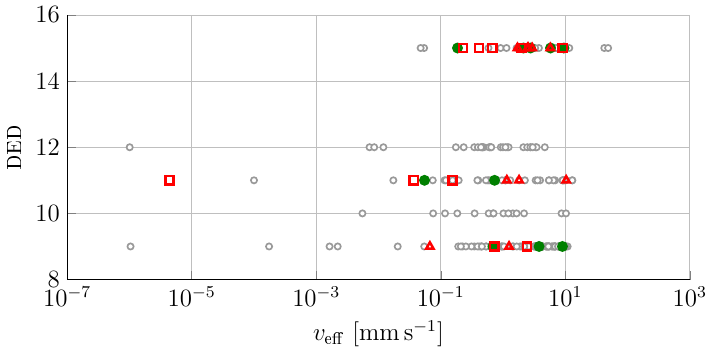}\vspace*{-0.5cm}\caption*{(f)}
    \end{minipage}
    \caption{\small \small Parameters estimated via fitting using the parametric models introduced in Section~\ref{sec:est}. Estimated parameters with \ac{RMSE} above threshold $\tau_n (15\%)$ are shown by gray circles. Parameters estimated by fitting the basic model $\hat{I}_{\mathrm{dist},1}(\varcal{P}^{1}, t)$ to the measurements with RMSEs below $\tau_1 (15\%)$ are shown by green markers. Pairs of estimated parameters obtained via fitting the extended model $\hat{I}_{\mathrm{dist},2}(\varcal{P}^{2}, t)$ to the measurements with RMSEs below $\tau_2 (15\%)$ are shown by red triangles and rectangles. }
    \label{fig:Params}
\end{figure*}

Figure~\ref{fig:Params} shows the parameters estimated via curve fitting for the basic and extended model, in six sub-figures. Each sub-figure visualizes pairwise correlations within the parameter set $\{d_{\mathrm{inj}}, \Leff, \Deff, d_{\mathrm{rx}}, \text{DED}, \veff\}$. Here, $\Leff, \Deff, d_{\mathrm{rx}}$, and $\veff$ are obtained via curve fitting, whereas $\text{DED}$ and $d_{\mathrm{inj}}$ are measured. The gray circles indicate the estimated parameters corresponding to approximations with RMSE values above $\tau_n(15\%)$, i.e., $\{\varcal{P}^n|\text{RMSE}(\hat{I}_{\mathrm{dist},n})>\tau_n(15\%)\}$. The green markers show parameters estimated based on the basic model $\hat{I}_{\mathrm{dist},1}(\varcal{P}^{1},t)$ with \ac{RMSE}s below $\tau_1(15\%)$, i.e., $\{\varcal{P}^{1}|\text{RMSE}(\hat{I}_{\mathrm{dist},1})<\tau_1(15\%)\}$. 
For the extended model $\hat{I}_{\mathrm{dist},2}(\varcal{P}^{2},t)$, there are pairs for each estimated parameter as two sets are considered for $n=2$. Therefore, the red triangle and rectangle markers in Figure~\ref{fig:Params} denote the estimated parameters belonging to parameter sets $\mathcal{P}^{2,1}$ and $\mathcal{P}^{2,2}$ for the extended parametric model $\hat{I}_{\mathrm{dist},2}(\varcal{P}^{2},t)$, i.e., $\{\varcal{P}^{2,1}|\text{RMSE}(\hat{I}_{\mathrm{dist},2})<\tau_2(15\%)\}$ and $\{\varcal{P}^{2,2}|\text{RMSE}(\hat{I}_{\mathrm{dist},2})<\tau_2(15\%)\}$. 

Figure~\ref{fig:Params}~(a) shows the estimated effective length $\Leff$ on the horizontal axis and the measured injection distance, $d_{\mathrm{inj}}$, on the vertical axis. It can be observed that the estimated $\Leff$ values are in the same order of magnitude as the distance measured in the \ac{CAM} model. This observation suggests that the actual range of the channel length is $2.5-5~\si{\centi\meter}$. 
Figure~\ref{fig:Params}~(b) shows the correlation between the estimated $\Leff$ and $\Deff$ values. First, it can be observed that the estimated $\Deff$ values with \ac{RMSE} $<\tau_n(15\%)$ are consistently within the range $10^{-6}-10^{-5}~\si{\square\meter\per\second}$. Moreover, we can observe that $\Deff$ increases from $10^{-7}~\si{\square\meter\per\second}$ to $10^{-5}~\si{\square\meter\per\second}$, when the estimated value of $\Leff$ increases from $0.5~\si{\centi\meter}$ to $7~\si{\centi\meter}$. 
This observation indicates that larger $\Deff$ (more dispersion) values are often obtained jointly with larger $\Leff$ values. This is directly correlated to the dynamics of the measurements shown in Fig.~\ref{fig:2Ring}, i.e., when the measurements are more dispersive (smoother increase and fewer peaks), the models account for this behavior by larger $\Deff$ and $\Leff$ values. For example, the dynamics of the measured particle distribution in Egg 1 (top left in Fig.~\ref{fig:2Ring}) are much smoother than those of Egg 20 (bottom left in Fig.~\ref{fig:2Ring}), and therefore, the estimated $\Deff$ and $\Leff$ values are larger for Egg 1 compared to Egg 20. 
Figure~\ref{fig:Params}~(c) shows the injection distance measured for the \ac{CAM} model, $d_{\mathrm{inj}}$, between the injection point and the \ac{ROI} located at $\bm{x}_\mathcal{R}$ over the estimated $d_{\mathrm{rx}}$ values. We observe that the estimated values of the observation point, $d_\mathrm{rx}$, are in agreement with $d_{\mathrm{inj}}$. This observation confirms that the estimates obtained for $d_{\mathrm{rx}}$ are in a reasonable range. Inspecting the gray line (i.e., $d_\mathrm{rx} = d_{\mathrm{inj}}$) in Fig.~\ref{fig:Params}~(c), it can be observed that most of the estimated $d_{\mathrm{rx}}$ values with \ac{RMSE} $<\tau_n(15\%)$ are below this line, indicating that the estimated $d_\mathrm{rx}$ is larger than $d_{\mathrm{inj}}$. This observation is reasonable because most of the vessels are curved and obviously the length of the vessel between the injection point and the \ac{ROI} is larger than the straight line connecting both points.
Figure~\ref{fig:Params}~(d) shows the \ac{DED} on the vertical axis and the estimated $\Deff$ values on the horizontal axis. Inspecting only the estimated $\Deff$ values with \ac{RMSE} $<\tau_n(15\%)$, it can be observed that the variance of the estimated values decreases with increasing \ac{DED}.
One reason for this could be that eggs at higher \acp{DED}, i.e., more developed and older eggs, have larger vessels that mainly dominate the particle distribution processes. Consequently, the transport dynamics and physical channel in those eggs more closely align with the assumptions of our analytical model, which is based on molecular advection-diffusion in a closed-loop system\footnote{This observation suggests that more reliable estimates can be obtained in older eggs. Therefore, we will investigate the correlation between the estimated parameters and the \ac{DED} further by extending our experimental studies to older eggs in future work.}.
Figures~\ref{fig:Params}~(e) and (f) show the relationship between the \ac{DED} and the estimated $\Leff$ and $\veff$ values, respectively. First, for both parameters, it can be observed that the variance of the estimated values also decreases as the \ac{DED} increases.
Moreover, from Fig.~\ref{fig:Params}~(f) it can be observed that the estimated $\veff$ values are consistently in the range from $10^{-2}-10~\si{\milli\meter\per\second}$, which is also within the range expected from the literature for velocities inside the \ac{CAM} vascular system (see Section~\ref{sec:sec2B}).

Table~\ref{tab:datas} shows the mean values for each estimated parameter for the parameter sets shown in Fig.~\ref{fig:Params}.
From Table~\ref{tab:datas} it can be observed that the mean values of the estimated parameters cluster around similar values within the dataset. For example, the mean of the estimated $\Deff$ values is in the same order of magnitude for all considered parameter sets, including those with \ac{RMSE} $>\tau(15\%)$.
This observation reveals a consistent trend across different models. Furthermore, the parameters seem to be representative of the CAM model, making them suitable for future theoretical research particularly in characterizing closed-loop channels inspired by the CAM model. The analysis of the parametric models for the distribution of molecules in the \ac{CAM} during the \textit{transient} and \textit{steady state phases} showed that both the basic model in \eqref{eq:approx} and the extended model in \eqref{eq:approx2} are capable of approximating the distribution dynamics while providing reasonable estimates of the physical parameters. 

\begin{table*}[t!]
\begin{center}
\caption{Mean of the estimated parameters shown in Fig.~\ref{fig:Params}.}
    \label{tab:datas}
\begin{tabular}{|c|c|c|c|c|}
\hline
\multirow{2}{*}{~} & \multicolumn{4}{|c|}{mean}\\
\cline{2-5}
 & $\Deff~\left[\si{\square\meter\per\second}\right]$ & $\veff~\left[\si{\milli\meter\per\second}\right]$ & $\Leff~\left[\si{\centi\meter}\right]$ & $d_\mathrm{rx}~\left[\si{\centi\meter}\right]$\\
\hline
$\{\mathcal{P}^n|\text{RMSE}(\hat{I}_{\mathrm{dist},n})>\tau_{n}(15\%)\}$        &$6.2\cdot 10^{-6}$&3.8&3.6&1.7\\
\hline
$\{\mathcal{P}^1|\text{RMSE}(\hat{I}_{\mathrm{dist},1})<\tau_{1}(15\%)\}$        &$5\cdot 10^{-6}$  &3.5&3.4&1.7\\
\hline
$\{\mathcal{P}^{2,1}|\text{RMSE}(\hat{I}_{\mathrm{dist},2})<\tau_{2}(15\%)\}$    &$5.7\cdot 10^{-6}$&3  &4  &2.2\\
\hline
$\{\mathcal{P}^{2,2}|\text{RMSE}(\hat{I}_{\mathrm{dist},2})<\tau_{2}(15\%)\}$    &$4.8\cdot 10^{-6}$&1.6&3.5&1.4\\
\hline
\end{tabular}
\end{center}
\end{table*}


\subsection{Accumulation Phase}
\label{sec:acc}
\subsubsection{Measurement Process}
As described in Section~\ref{sec:acc_est} the \textit{accumulation phase} starts after the \textit{steady state phase}, i.e, $t > t_\mathrm{acc}$. Hence, the measurement time was increased to approximately one hour. To overcome the challenge of possible bleeding after removing the injection needle, we fixed the syringe throughout the measurement time (see Fig.~\ref{fig:Liver}). We used $20$ eggs, but so far only two were successfully measurable. The percentage of failure is higher in this case because the liver has to remain visible for the whole observation time\footnote{The unsuccessful measurements (eggs) are not wasted as we used them for other experiments and data generation for the \textit{transient} and \textit{steady state phases}.}. However, due to the embryo's movement the liver position changes and may move under the egg shell or be blocked by other tissue. Since only two successful measurements could be obtained, they are not yet part of the published dataset. In the future, we will extend the dataset after obtaining a sufficient number of successful measurements for the molecule accumulation in the embryo's liver.

\subsubsection{Molecule Accumulation in the Liver}

\begin{figure}[t!]
    \centering
    \includegraphics[width=0.9\linewidth]{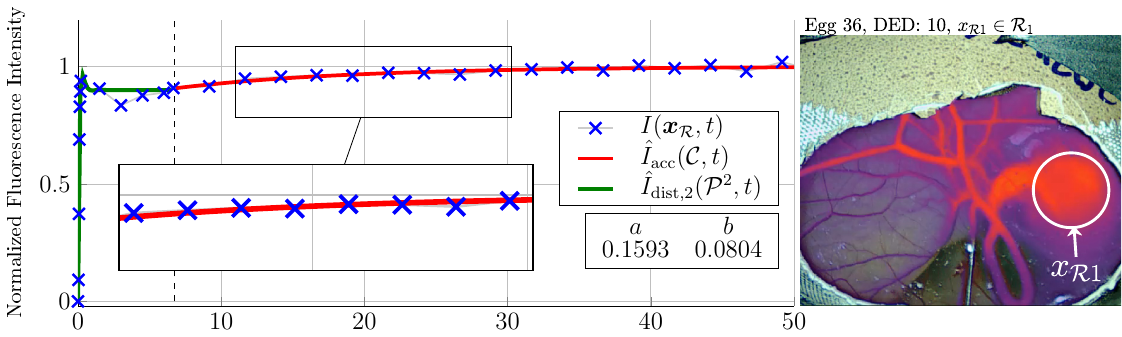}\\[-0.5em]
    \includegraphics[width=0.9\linewidth]{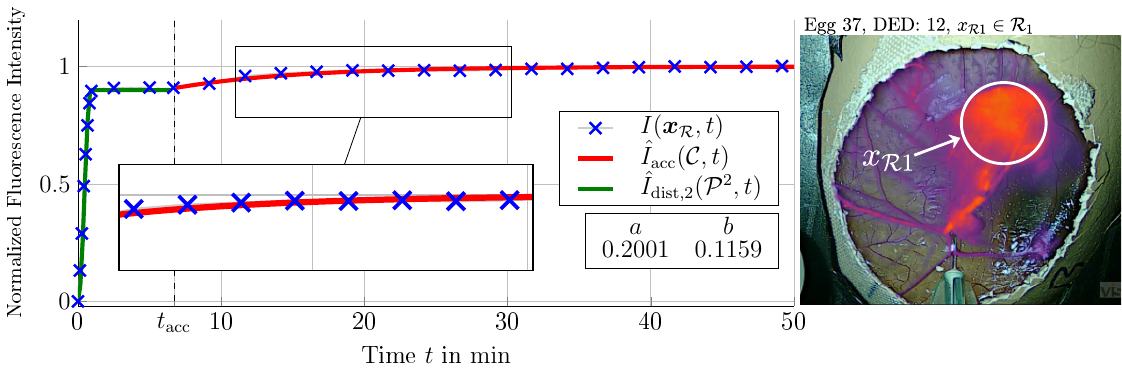}
    \caption{\small Experimental and fitting results of \ac{ICG} distribution in the embryo's liver. The \textit{transient} and \textit{steady state phases} are approximated by $\hat{I}_{\mathrm{dist},2}(\varcal{P}^2,t)$ based on~\eqref{eq:approx2}. The \textit{accumulation phase} is approximated by $\hat{I}_{\mathrm{acc}}(\varcal{C},t)$ based on~\eqref{eq:accum}. The blue markers correspond to the measured signal. The vertical dashed line indicates $t_\mathrm{acc}$. The photos on the right hand side show the injection point and the ROIs.}
    \label{fig:LiverHat}
\end{figure}

Figure~\ref{fig:LiverHat} shows the measured \ac{ICG} fluorescence intensity $I(\bm{x}_\mathcal{R},t)$ over time (curves with blue markers). The \ac{ROI} $\mathcal{R}$ on the right hand side of Fig.~\ref{fig:LiverHat} indicates the region where the liver of the embryo is located. The green solid curves corresponds to the approximation of the fluorescence intensity during the \textit{transient} and \textit{steady state phases}, obtained by fitting the extended parametric model \eqref{eq:approx2} to the measurement data. The red solid curves represent the approximation of the fluorescence intensity during the \textit{accumulation phase}, obtained by fitting $\hat{I}_{\mathrm{acc}}(\mathcal{C},t)$ in \eqref{eq:accum} to the measurement data. Moreover, the transition time between the \textit{steady state phase} and the \textit{accumulation phase} $t = t_\mathrm{acc}$ is shown by a vertical dashed line.
The results shown in Fig.~\ref{fig:LiverHat} confirm our modeling assumptions made in Section~\ref{subsec:assumpt}, i.e., that the molecule distribution process inside the \ac{CAM} model can be divided into three distinct phases. It can be observed that the \textit{accumulation phase} starts after the \textit{steady state phase}, where the fluorescence intensity in the region of the liver slowly starts to increase. Moreover, we can observe from Fig.~\ref{fig:LiverHat} that the proposed simple parametric model for the accumulation in~\eqref{eq:accum} is in very good agreement with the measurement data, which is confirmed by the \ac{RMSE}$\left(\hat{I}_\mathrm{acc}\right)$ values, i.e., $1\cdot 10^{-4}$ (Egg 36) and $3\cdot 10^{-5}$ (Egg 37).  
For both experiments, the parameter sets $\mathcal{C}$ estimated via the fitting process are in the same range, indicating the validity of the proposed accumulation model. 

\section{Discussion and Future Work}
\label{sec:discuss}

In the previous sections, we have shown that the proposed parametric models are capable of approximating the molecule distribution and accumulation within the closed-loop CAM vascular system. However, there are still
deviations from the measurement data, as the proposed models are not able to capture all environmental effects occurring in the CAM model.  
In the following, we present two lists, where the first one is a discussion of modeling inaccuracies and approaches to further enhance the modeling accuracy in future work. The second list describes the experimental challenges and provides methods to further improve the experimental setup.
\begin{itemize}
    \item The measurements of the fluorescence camera are 2D representations of a 3D system. Therefore, the measured fluorescence intensity at $\bm{x}_\varcal{R}$ can be interfered by the fluorescence of other vessels lying below the vessel of interest. Moreover, the measurements can be strongly influenced by random movements of the embryo. 
    \item The model does not account for chemical reactions that may occur between the \ac{ICG} molecules and other molecules in the environment. Moreover, the accumulation of \ac{ICG} in the yolk bag and the embryo’s organs and their binding to vessel walls need to be investigated. Specifically, the proposed model for the molecule accumulation in the liver should be extended to account for the actual binding and accumulation processes. 
    \item The effective flow velocity $\veff$ in \eqref{eq:wrappedNormal} is assumed to be constant. However, the blood flow velocity inside the \ac{CAM} model is not constant but time-variant, which has to be taken into account. 
\end{itemize}
Next, we highlight future research directions for improving the experimental setup.
\begin{itemize}
    \item One of the challenges in monitoring particle propagation inside the CAM model is the embryo's movement. In the future, we plan to minimize motion-induced artifacts. In particular, the embryo could be immobilized by placing the egg in ice chips for $20~\si{\minute}$ to $90~\si{\minute}$ or by dropping $5~\si{\milli\gram}$ Ketamine onto the CAM model surface~\cite{winter2020vivo}.

    \item Monitoring the CAM continuously for long intervals (e.g., days) is crucial for studying long term effects. However, we cannot keep the eggs outside the incubator for extended periods of time since the environment must maintain certain levels of temperature and humidity. Another reason is the immunodeficiency of the CAM, which makes it vulnerable to contamination. Hence, in the future, we plan to install our measurement setup inside an incubator.
    
    \item Repeatability of the experiments is crucial for more consistent and more interpretable results. Currently, the injection procedure is performed individually and by hand for every measurement. Therefore, to obtain repeatable measurements with high confidence, we plan to develop a setup with micro-injectors in the future.
    
    \item In this paper, we considered only one injection of a fixed amount of \ac{ICG} ($50~\si{\micro\liter}$) to obtain the ``impulse response'' of the system. In the future, we will extend our experiments to different amounts of \ac{ICG} to study the dependency on the injected \ac{ICG} concentration and to repeated injections within one measurement. 
\end{itemize}
\section{conclusion}
\label{sec:conc}
In this paper, we introduced the \ac{CAM} model as a versatile \textit{in vivo} \ac{MC} testbed as an important step to enable the transition from proof-of-concept \ac{MC} systems towards practical applications. The CAM model consists of a closed-loop cardiovascular system including blood circulation and organs, and therefore provides a realistic environment for the validation and optimization of MC technology.  
We introduced an analytical model for the diffusion and flow of molecules in dispersive closed-loop systems and validated it by \ac{PBS}. We experimentally investigated and measured the distribution of the fluorescent molecule \ac{ICG} in the \ac{CAM} model, and created and published a dataset of $69$ measurements in $25$ different eggs. From the measurements, we identified three different phases for the molecule distribution in the \ac{CAM} model, i.e., the \textit{transient phase}, \textit{steady state phase}, and \textit{accumulation phase}. The \textit{transient phase} corresponds to the quick distribution of the molecules in the \ac{CAM}'s vascular system. The \textit{steady state phase} starts after the \textit{transient phase}, where the \ac{ICG} concentration remains approximately constant. In the \textit{accumulation phase}, molecules start to leave the vascular system and accumulate in the liver of the chick embryo.
Based on the derived analytical model, we proposed two parametric models to approximate the distribution of \ac{ICG} molecules in the CAM model during the \textit{transient} and \textit{steady state phases}. Our results showed that the proposed parametric models are capable of accurately approximating the molecule distribution inside the CAM model. Moreover, we investigated the overall approximation performance of the proposed models in terms of the \ac{RMSE}. Our results revealed that the estimated parameters that were obtained by fitting the parametric models to the measurements are consistent and partly in a plausible range compared to empirically observed parameters. Furthermore, we proposed a simple parametric model to approximate the accumulation of molecules in the liver of the embryo during the \textit{accumulation phase}. Based on a small experimental study, we demonstrated that the proposed model successfully captures the effects occurring in the \textit{accumulation phase}. 
Finally, we discussed challenges and outlined future research directions, where the focus was particularly placed on the extension of the developed analytical models to account for more relevant environmental effects inside the \ac{CAM} model, e.g., chemical reactions, and on the improvement of the testbed setup. 

\appendices
\section{Methods}\label{sec:methods}
For all experiments, chicken eggs were incubated at a constant temperature of $37.8~\si{\celsius}$, a pCO$_2$ of $5~\si{\percent}$, and humidity calibrated to $63~\si{\percent}$. The eggs were rotated during the first four days of incubation to prevent the embryos from adhering to the shell membranes and to minimize the risk of vascular damage when opening the eggshells. The shells were opened on \ac{DED} 4, before the \ac{CAM} fully fused to the shell membranes, to avoid damaging the developing vasculature \cite{nowak2014chicken}.
To this end, the air entrapment in the egg was localized with a light source and a hole of approximately $2~\si{\milli\meter} \times 2~\si{\milli\meter}$ was cut with sterile tweeters into the overlaying eggshell to introduce atmospheric negative pressure. A second hole of approximately $5~\si{\milli\meter} \times 5~\si{\milli\meter}$ was introduced at the longitudinal side of the egg. The holes were covered with Leukosilk® (BSN medical, Hamburg, Germany) and the eggs were returned to the incubator in a static position. ICG was administered intravenously to assess intratumoral perfusion. The ICG solution was prepared by diluting it to a concentration of $50~\si{\mu\molar}$ in $0.9~\si{\percent}$ sodium chloride, resulting in a final dosage of $0.3~\si{\milli\gram\per\kilo\gram}$, additionally 5~\si{\percent} blood was added to the solution. Fluorescence peaked at a wavelength of $830~\si{\nano\meter}$. Imaging was conducted using the Medtronic Elevision IR (VSIII) fluorescence system (Medtronic®, Meerbusch, Germany), which operates in the near-infrared spectrum. The injection process employed a $1~\si{\milli\liter}$ syringe (Norm-Ject HENKE, Tuttlingen, Germany) fitted with a 33~G needle (Hypodermic Needles, 33 G 0.20~$\si{\milli\meter}$ × 4~$\si{\milli\meter}$, MESORAM, Putzbrunn, Germany).
The video recordings of the \ac{ICG} molecule distribution were performed with an \ac{ICG} camera (EleVision\textsuperscript{TM} IR Platform, Medtronic).
The video data analysis (Fluorescence tracking in \acp{ROI}) was done with the Matlab Fluorescence Tracker App\footnote{\url{https://matlab.mathworks.com/open/github/v1?repo=mathworks/Fluorescence-Tracker-App}}.
We measured the fluorescent intensity of pure ICG beside the egg and used it as a reference signal. This reference signal was then subtracted from the fluorescent intensity measured in the desired ROI. This subtraction was essential to account for variations in ICG fluorescence caused by the inherent properties of the ICG molecule and automatic light intensity rescaling performed by the camera. The sampling time step $\Delta t$ was $10^{-4}~\si{\second}$ and $10^{-2}~\si{\second}$ for $I_{\mathrm{dist}}(\bm{x}_{\varcal{R}}, t)$ and $I_{\mathrm{acc}}(\bm{x}_{\varcal{R}}, t)$, respectively. 
\bibliographystyle{IEEEtran}
\bibliography{IEEEabrv,./literature.bib}

\begin{thebibliography}{10}
\providecommand{\url}[1]{#1}
\csname url@samestyle\endcsname
\providecommand{\newblock}{\relax}
\providecommand{\bibinfo}[2]{#2}
\providecommand{\BIBentrySTDinterwordspacing}{\spaceskip=0pt\relax}
\providecommand{\BIBentryALTinterwordstretchfactor}{4}
\providecommand{\BIBentryALTinterwordspacing}{\spaceskip=\fontdimen2\font plus
\BIBentryALTinterwordstretchfactor\fontdimen3\font minus \fontdimen4\font\relax}
\providecommand{\BIBforeignlanguage}[2]{{%
\expandafter\ifx\csname l@#1\endcsname\relax
\typeout{** WARNING: IEEEtran.bst: No hyphenation pattern has been}%
\typeout{** loaded for the language `#1'. Using the pattern for}%
\typeout{** the default language instead.}%
\else
\language=\csname l@#1\endcsname
\fi
#2}}
\providecommand{\BIBdecl}{\relax}
\BIBdecl

\bibitem{schafer2024chorioallantoic}
M.~Sch{\"a}fer \emph{et~al.}, ``The chorioallantoic membrane model: A {3D} in vivo testbed for design and analysis of {MC} systems,'' in \emph{Proc. 11th Ann. ACM Int. Conf. Nanoscale Comp. Comm.}, 2024, pp. 47--53.

\bibitem{Nakano2011}
T.~Nakano, A.~W. Eckford, and T.~Haraguchi, \emph{Molecular Communication}.\hskip 1em plus 0.5em minus 0.4em\relax Cambridge University Press, 2011.

\bibitem{Farsad2016}
N.~Farsad, H.~B. Yilmaz, A.~Eckford, C.~Chae, and W.~Guo, ``A comprehensive survey of recent advancements in molecular communication,'' \emph{IEEE Commun. Surv. Tutor.}, vol.~18, no.~3, pp. 1887--1919, 2016.

\bibitem{Felicetti2016}
L.~Felicetti, M.~Femminella, G.~Reali, and P.~Liò, ``Applications of molecular communications to medicine: {A} survey,'' \emph{Nano Commun. Netw.}, vol.~7, pp. 27--45, 2016.

\bibitem{Lotter2023b}
S.~Lotter \emph{et~al.}, ``Experimental research in synthetic molecular communications – {P}art {I},'' \emph{IEEE Nanotechnol. Mag.}, vol.~17, no.~3, pp. 42--53, 2023.

\bibitem{Farsad2017}
N.~Farsad, D.~Pan, and A.~Goldsmith, ``A novel experimental platform for in-vessel multi-chemical molecular communications,'' in \emph{Proc. IEEE Global Commun. Conf.}, 2017, pp. 1--6.

\bibitem{Unterweger2018}
H.~Unterweger \emph{et~al.}, ``Experimental molecular communication testbed based on magnetic nanoparticles in duct flow,'' in \emph{IEEE 19th Int. Workshop Signal Process. Adv. Wireless Commun.}, 2018, pp. 1--5.

\bibitem{cali2024experimental}
F.~Cal{\`\i} \emph{et~al.}, ``Experimental implementation of molecule shift keying for enhanced molecular communication,'' \emph{IEEE Trans. Mol. Biol. Multi-Scale Commun}, vol.~10, no.~1, pp. 175--184, 2024.

\bibitem{Pan2022}
W.~o. Pan, ``A molecular communication platform based on body area nanonetwork,'' \emph{Nanomater.}, vol.~12, no.~4, 2022.

\bibitem{Wang2020}
J.~Wang, D.~Hu, C.~Shetty, and H.~Hassanieh, ``Understanding and embracing the complexities of the molecular communication channel in liquids,'' in \emph{Proc. 26th Ann. Int. Conf. Mobile Comp. Netw.}, 2020, pp. 1--15.

\bibitem{huang2024energy}
Y.~Huang \emph{et~al.}, ``An energy-efficient ternary modulation with water for molecular communication systems: From solvent to information carrier,'' \emph{IEEE Trans. Mol. Biol. Multi-Scale Commun}, vol.~10, no.~2, pp. 236--242, 2024.

\bibitem{lin2024testbed}
L.~Lin, W.~Wang, W.~Yu, and H.~Yan, ``Testbed for molecular communication system based on light absorption: Study of information transmission from inside to outside body,'' \emph{IEEE Trans. Mol. Biol. Multi-Scale Commun}, vol.~10, no.~2, pp. 212--222, 2024.

\bibitem{Brand2022}
L.~Brand \emph{et~al.}, ``Media modulation based molecular communication,'' \emph{IEEE Trans. Commun.}, vol.~70, no.~11, pp. 7207--7223, 2022.

\bibitem{Brand2024}
------, ``Switchable signaling molecules for media modulation: Fundamentals, applications, and research directions,'' \emph{IEEE Commun. Mag.}, vol.~62, no.~5, pp. 112--118, 2024.

\bibitem{Brand2024closed}
------, ``Closed loop molecular communication testbed: Setup, interference analysis, and experimental results,'' in \emph{Proc. IEEE Int. Conf. Commun.}, 2024, pp. 4805--4811.

\bibitem{Lotter2023a}
S.~Lotter \emph{et~al.}, ``Experimental research in synthetic molecular communications – part {II},'' \emph{IEEE Nanotechnol. Mag.}, vol.~17, no.~3, pp. 54--65, 2023.

\bibitem{gilbert2011developmental}
S.~F. Gilbert, \emph{\BIBforeignlanguage{English (US)}{Developmental {B}iology}}.\hskip 1em plus 0.5em minus 0.4em\relax Sinauer Associates, Sunderland, Mass., 2003.

\bibitem{Mapanao21}
A.~K. Mapanao \emph{et~al.}, ``Tumor grafted – chick chorioallantoic membrane as an alternative model for biological cancer research and conventional/nanomaterial-based theranostics evaluation,'' \emph{Expert Opin. Drug Metab. Toxicol.}, vol.~17, no.~8, pp. 947--968, 2021.

\bibitem{Valdes2002}
T.~Valdes, D.~Kreutzer, and F.~Moussy, ``The chick chorioallantoic membrane as a novel in vivo model for the testing of biomaterials,'' \emph{J. Biomed. Mater. Res.}, vol.~62, pp. 273--282, 2002.

\bibitem{AngioRef}
G.~Santulli, \emph{\BIBforeignlanguage{English (US)}{Angiogenesis: Insights From a Systematic Overview}}.\hskip 1em plus 0.5em minus 0.4em\relax Nova Science Publishers, Inc., 2013.

\bibitem{Feder2020}
A.~L. Feder \emph{et~al.}, ``Extended analysis of intratumoral heterogeneity of primary osteosarcoma tissue using 3{D}-in-vivo-tumor-model,'' \emph{Clin. Hemorheol. Microcirc.}, vol.~76, no.~2, pp. 133--141, 2020.

\bibitem{kohl20223d}
C.~Kohl \emph{et~al.}, ``The {3D} in vivo chorioallantoic membrane model and its role in breast cancer research,'' \emph{J. Cancer Res. Clin. Oncol.}, vol. 148, no.~5, pp. 1033--1043, 2022.

\bibitem{Wagner24}
B.~J. Wagner \emph{et~al.}, ``Patient-derived xenografts from circulating cancer stem cells as a preclinical model for personalized pancreatic cancer research,'' \emph{Sci. Rep.}, vol.~15, no.~1, p. 2896, 2025.

\bibitem{Ettner2024}
A.~Ettner-Sitter \emph{et~al.}, ``Visualization of vascular perfusion of human pancreatic cancer tissue in the {CAM} model and its impact on future personalized drug testing,'' \emph{Organoids}, vol.~3, no.~1, pp. 1--17, 2024.

\bibitem{ICGdistDataset2025}
\BIBentryALTinterwordspacing
F.~Vakilipoor \emph{et~al.}, ``Transient indocyanine green distribution measurement and modelling in chorioallantoic membrane ({CAM}) model,'' Mar. 2025. [Online]. Available: \url{https://doi.org/10.5281/zenodo.14626107}
\BIBentrySTDinterwordspacing

\bibitem{rous1911tumor}
P.~Rous, ``Tumor implantations in the developing embryo. experiments with a transmissible sarcoma of the fowl,'' \emph{J. Amer. Med. Assoc.}, vol.~56, pp. 741--742, 1911.

\bibitem{murphy1912transplantability}
J.~B. Murphy, ``Transplantability of malignant tumors to the embryos of a foreign species,'' \emph{J. Amer. Med. Assoc.}, vol.~59, no.~11, pp. 874--875, 1912.

\bibitem{goodpasture1931cultivation}
E.~W. Goodpasture, A.~M. Woodruff, and G.~Buddingh, ``The cultivation of vaccine and other viruses in the chorio-allantoic membrane of chick embryos,'' \emph{Science}, vol.~74, no. 1919, pp. 371--372, 1931.

\bibitem{moore1942chorioallantoic}
M.~Moore, ``The chorioallantoic membrane of chick embryos and its response to inoculation with some mycobacteria,'' \emph{Amer. J. Pathol.}, vol.~18, no.~5, p. 827, 1942.

\bibitem{maibier2016structure}
M.~Maibier \emph{et~al.}, ``Structure and hemodynamics of vascular networks in the chorioallantoic membrane of the chicken,'' \emph{Amer. J. Physiol. Heart Circ. Physiol.}, vol. 311, no.~4, pp. H913--H926, 2016.

\bibitem{defouw1989mapping}
D.~O. DeFouw, V.~J. Rizzo, R.~Steinfeld, and R.~N. Feinberg, ``Mapping of the microcirculation in the chick chorioallantoic membrane during normal angiogenesis,'' \emph{Microvasc. Res.}, vol.~38, no.~2, pp. 136--147, 1989.

\bibitem{shumko1988vascular}
J.~Z. Shumko, D.~O. Defouw, and R.~N. Feinberg, ``Vascular histodifferentiation in the chick chorioallantoic membrane: a morphometric study,'' \emph{Anat. Rec.}, vol. 220, no.~2, pp. 179--189, 1988.

\bibitem{nikiforidis1999quantitative}
G.~Nikiforidis \emph{et~al.}, ``Quantitative assessment of angiogenesis in the chick embryo and its chorioallantoic membrane by computerised analysis of angiographic images,'' \emph{Eur. J. Radiol.}, vol.~29, no.~2, pp. 168--179, 1999.

\bibitem{smith2016microvascular}
A.~F. Smith \emph{et~al.}, ``Microvascular hemodynamics in the chick chorioallantoic membrane,'' \emph{Microcirc.}, vol.~23, no.~7, pp. 512--522, 2016.

\bibitem{staton2004current}
C.~A. Staton \emph{et~al.}, ``Current methods for assaying angiogenesis in vitro and in vivo,'' \emph{Int. J. Exp. Pathol.}, vol.~85, no.~5, pp. 233--248, 2004.

\bibitem{storgard2005angiogenesis}
C.~Storgard, D.~Mikolon, and D.~G. Stupack, ``Angiogenesis assays in the chick {CAM},'' \emph{Cell Migr. Dev. Methods Protoc.}, pp. 123--136, 2005.

\bibitem{ribatti2008chick}
D.~Ribatti, ``The chick embryo chorioallantoic membrane in the study of tumor angiogenesis,'' \emph{Rom. J. Morphol. Embryol.}, vol.~49, no.~2, pp. 131--135, 2008.

\bibitem{mesas2024experimental}
C.~Mesas \emph{et~al.}, ``Experimental tumor induction and evaluation of its treatment in the chicken embryo chorioallantoic membrane model: A systematic review,'' \emph{Int. J. Mol. Sci.}, vol.~25, no.~2, p. 837, 2024.

\bibitem{chase2017development}
D.~M. Chase, D.~J. Chaplin, and B.~J. Monk, ``The development and use of vascular targeted therapy in ovarian cancer,'' \emph{Gynecol. Oncol.}, vol. 145, no.~2, pp. 393--406, 2017.

\bibitem{chu2022applications}
P.-Y. Chu, A.~P.-F. Koh, J.~Antony, and R.~Y.-J. Huang, ``Applications of the chick chorioallantoic membrane as an alternative model for cancer studies,'' \emph{Cells Tissues Organs}, vol. 211, no.~2, pp. 222--237, 2022.

\bibitem{chen2021utilisation}
L.~Chen \emph{et~al.}, ``Utilisation of chick embryo chorioallantoic membrane as a model platform for imaging-navigated biomedical research,'' \emph{Cells}, vol.~10, no.~2, p. 463, 2021.

\bibitem{zhou2013novel}
Q.~Zhou \emph{et~al.}, ``A novel four-step system for screening angiogenesis inhibitors,'' \emph{Mol. Med. Rep.}, vol.~8, no.~6, pp. 1734--1740, 2013.

\bibitem{dunker2019implementation}
N.~D{\"u}nker and V.~Jendrossek, ``Implementation of the chick chorioallantoic membrane ({CAM}) model in radiation biology and experimental radiation oncology research,'' \emph{Cancers}, vol.~11, no.~10, p. 1499, 2019.

\bibitem{palumbo2023cam}
C.~Palumbo, F.~Sisi, and M.~Checchi, ``{CAM} model: Intriguing natural bioreactor for sustainable research and reliable/versatile testing,'' \emph{Biology}, vol.~12, no.~9, p. 1219, 2023.

\bibitem{nowak2014chicken}
P.~Nowak-Sliwinska, T.~Segura, and M.~L. Iruela-Arispe, ``The chicken chorioallantoic membrane model in biology, medicine and bioengineering,'' \emph{Angiogenesis}, vol.~17, pp. 779--804, 2014.

\bibitem{pawlikowska2020exploitation}
P.~Pawlikowska \emph{et~al.}, ``Exploitation of the chick embryo chorioallantoic membrane ({CAM}) as a platform for anti-metastatic drug testing,'' \emph{Sci. Rep.}, vol.~10, no.~1, p. 16876, 2020.

\bibitem{lange2001new}
N.~Lange, J.-P. Ballini, G.~Wagnieres, and H.~van~den Bergh, ``A new drug-screening procedure for photosensitizing agents used in photodynamic therapy for {CNV},'' \emph{Invest. Ophthalmol. Vis. Sci.}, vol.~42, no.~1, pp. 38--46, 2001.

\bibitem{li2011dual}
J.~Li \emph{et~al.}, ``A dual-targeting anticancer approach: soil and seed principle,'' \emph{Radiol.}, vol. 260, no.~3, pp. 799--807, 2011.

\bibitem{eichhorst2024establishment}
L.~Eichhorst \emph{et~al.}, ``Establishment of head and neck cancer patient-derived xenografts {PDX} in the chorion-allantois membrane assay ({CAM} assay),'' \emph{Laryngo-Rhino-Otologie}, vol. 103, no. S 02, 2024.

\bibitem{schueler2024ultra}
J.~Schueler \emph{et~al.}, ``Ultra high frequency ultrasound enables real-time visualization of blood supply from chorioallantoic membrane to human autosomal dominant polycystic kidney tissue,'' \emph{Sci. Rep.}, vol.~14, no.~1, p. 10063, 2024.

\bibitem{mcdonald2003imaging}
D.~M. McDonald and P.~L. Choyke, ``Imaging of angiogenesis: from microscope to clinic,'' \emph{Nat. Med.}, vol.~9, no.~6, pp. 713--725, 2003.

\bibitem{geng2018hatching}
L.~Geng \emph{et~al.}, ``Hatching eggs classification based on deep learning,'' \emph{Multimed. Tools Appl.}, vol.~77, pp. 22\,071--22\,082, 2018.

\bibitem{verhoelst2011structural}
E.~Verhoelst, ``Structural analysis of the angiogenesis in the chicken chorioallantoic membrane,'' Ph.D. dissertation, Katholieke Universiteit Leuven, 2011.

\bibitem{kundekova2021chorioallantoic}
B.~Kundekov{\'a} \emph{et~al.}, ``Chorioallantoic membrane models of various avian species: {D}ifferences and applications,'' \emph{Biology}, vol.~10, no.~4, p. 301, 2021.

\bibitem{Makanya2016}
A.~N. Makanya \emph{et~al.}, ``Dynamics of the developing chick chorioallantoic membrane assessed by stereology, allometry, immunohistochemistry and molecular analysis,'' \emph{PLoS One}, vol.~11, no.~4, pp. 1--23, 2016.

\bibitem{Guerra2021}
A.~Guerra \emph{et~al.}, ``Simulation of the process of angiogenesis: Quantification and assessment of vascular patterning in the chicken chorioallantoic membrane,'' \emph{Comput. Biol. Med.}, vol. 136, p. 104647, 2021.

\bibitem{Tazawa1977}
H.~Tazawa and M.~Mochizuki, ``Oxygen analyses of chicken embryo blood,'' \emph{Respir. Physiol.}, vol.~31, pp. 203--2015, 1977.

\bibitem{Kloosterman2014}
A.~Kloosterman, B.~Hierck, J.~Westerweel, and C.~Poelma, ``Quantification of blood flow and topology in developing vascular networks,'' \emph{PLoS One}, vol.~9, no.~5, pp. 1--13, 2014.

\bibitem{Maibier2016}
M.~Maibier \emph{et~al.}, ``Structure and hemodynamics of vascular networks in the chorioallantoic membrane of the chicken,'' \emph{Amer. J. of Physiol. Heart Circ. Physiol.}, vol. 311, no.~4, pp. H913--H926, 2016.

\bibitem{Weixler23}
B.~Weixler \emph{et~al.}, ``The value of indocyanine green image-guided surgery in patients with primary liver tumors and liver metastases,'' \emph{Life}, vol.~13, no.~6, 2023.

\bibitem{xiao:biol:2024}
X.~Xiao, D.~Yuan, Y.-X. Wang, and X.-A. Zhan, ``The protective effects of different sources of maternal selenium on oxidative stressed chick embryo liver,'' \emph{Biol. Trace Elem. Research}, vol. 172, no.~1, pp. 201--208, 2016.

\bibitem{yaseen2008vivo}
M.~A. Yaseen, J.~Yu, M.~S. Wong, and B.~Anvari, ``In-vivo fluorescence imaging of mammalian organs using charge-assembled mesocapsule constructs containing indocyanine green,'' \emph{Opt. Express}, vol.~16, no.~25, pp. 20\,577--20\,587, 2008.

\bibitem{Givisiez2020}
P.~E. Givisiez, A.~L. {Moreira Filho}, M.~R. Santos, H.~B. Oliveira, P.~R. Ferket, C.~J. Oliveira, and R.~D. Malheiros, ``Chicken embryo development: metabolic and morphological basis for in ovo feeding technology,'' \emph{Poultry Sci}, vol.~99, no.~12, pp. 6774--6782, 2020.

\bibitem{pion20223d}
E.~Pion \emph{et~al.}, ``3{D} in vivo models for translational research on pancreatic cancer: The chorioallantoic membrane ({CAM}) model,'' \emph{Cancers}, vol.~14, no.~15, p. 3733, 2022.

\bibitem{Miebach22}
L.~Miebach, J.~Berner, and S.~Bekeschus, ``In ovo model in cancer research and tumor immunology,'' \emph{Front. Immunol.}, vol.~13, 2022.

\bibitem{Zucal21}
I.~Zucal \emph{et~al.}, ``Indocyanine green for leakage control in isolated limb perfusion,'' \emph{J. Pers. Med.}, vol.~11, no.~11, 2021.

\bibitem{Zucal22}
------, ``An innovative simulation model for microvascular training,'' \emph{Plast. Reconstr. Surg.}, vol. 150, no.~1, pp. 189e--193e, 2022.

\bibitem{Chauhan23}
N.~Chauhan \emph{et~al.}, ``Indocyanine green-based glow nanoparticles probe for cancer imaging,'' \emph{Nanotheranostics}, vol.~7, no.~4, pp. 353--367, 2023.

\bibitem{Chon2023}
B.~Chon \emph{et~al.}, ``Indocyanine green ({ICG}) fluorescence is dependent on monomer with planar and twisted structures and inhibited by {H}-aggregation,'' \emph{Int. J. Mol. Sci.}, vol.~24, no.~17, 2023.

\bibitem{Jamali2019}
V.~Jamali \emph{et~al.}, ``Channel modeling for diffusive molecular communication— {A} tutorial review,'' \emph{Proc. IEEE}, vol. 107, no.~7, pp. 1256--1301, 2019.

\bibitem{Probstein2005}
R.~F. Probstein, \emph{Physiochemical Hydrodynamics: An Introduction}.\hskip 1em plus 0.5em minus 0.4em\relax John Wiley \& Sons, 2005.

\bibitem{Wicke2018}
W.~Wicke \emph{et~al.}, ``Modeling duct flow for molecular communications,'' in \emph{Proc. IEEE Global Commun. Conf.}, 2018, pp. 206--212.

\bibitem{Mardia1999}
K.~V. Mardia and P.~E. Jupp, \emph{Directional Statistics}.\hskip 1em plus 0.5em minus 0.4em\relax John Wiley \& Sons, 1999.

\bibitem{Schaefer2021}
M.~Schäfer \emph{et~al.}, ``Transfer function models for cylindrical {MC} channels with diffusion and laminar flow,'' \emph{IEEE Trans. Mol. Biol. Multi-Scale Commun.}, vol.~7, no.~4, pp. 271--287, 2021.

\bibitem{winter2020vivo}
G.~Winter \emph{et~al.}, ``In vivo {PET/MRI} imaging of the chorioallantoic membrane,'' \emph{Front. Phys.}, vol.~8, p. 151, 2020.

\end{thebibliography}


\end{document}